\documentclass[pdflatex,sn-mathphys]{sn-jnl}

\usepackage{graphicx}
\usepackage{multirow}
\usepackage{shadow}
\newcommand{\myhiding}[1]{}

\newif\ifcommentson\commentsonfalse
\newif\ifconferenceon\conferenceonfalse
\ifconferenceon
\newcommand{\arxiv}[1]{}
\newcommand{\conference}[1]{#1}
\newcommand{\conferenceShort}[1]{}
\else
\newcommand{\arxiv}[1]{#1}
\newcommand{\conference}[1]{}
\newcommand{\conferenceShort}[1]{}
\fi
\newcommand{\commentY}[1]{}

\jyear{2023}%

\theoremstyle{thmstyleone}%
\theoremstyle{thmstyletwo}%

\theoremstyle{thmstylethree}%

\newcommand\Sec[1] {Section~\ref{#1}}
\newcommand\Sects[1] {Sections~\ref{#1}}

\newcommand\Tbl[1] {Table~\ref{#1}}
\newcommand\Tbls[1] {Tables~\ref{#1}}
\newcommand\Fig[1] {Figure~\ref{#1}}

\newcommand{\spB}{{\hspace{-2ex}}}

\newcommand{\textgt}[1]{{\bf #1}}

\newcommand{\Rare}[0]{$\!\blacktriangle$}

\allowdisplaybreaks[1]

\usepackage{array}
\makeatletter
\newdimen\arrayrulewidthb
\arrayrulewidthb=\p@\relax
\newcolumntype{I}{!{\vrule width \arrayrulewidthb}}
\def\hlineb{%
  \noalign{\ifnum0=`}\fi\hrule \@height \arrayrulewidthb \futurelet
   \reserved@a\@xhlineb}
\def\@xhlineb{\ifx\reserved@a\hlineb
               \vskip\doublerulesep
               \vskip-\arrayrulewidthb
             \fi
      \ifnum0=`{\fi}}
\makeatother

\begin{document}

\title[Threats, Vulnerabilities, and Controls of Machine Learning Based Systems]{Threats, Vulnerabilities, and Controls of Machine Learning Based Systems: A Survey and Taxonomy}

\author*[1]{\fnm{Yusuke} \sur{Kawamoto}}

\author[1,2]{\fnm{Kazumasa} \sur{Miyake}}

\author[1]{\fnm{Koichi} \sur{Konishi}}

\author[1]{\fnm{Yutaka} \sur{Oiwa}}

\affil*[1]{\orgname{National Institute of Advanced Industrial Science and Technology (AIST)}, \orgaddress{\city{Tokyo}, \country{Japan}}}

\affil[2]{\orgname{Sumitomo Electric Industries, Ltd.}, \orgaddress{\city{Yokohama}, \country{Japan}}.~\\~\\~}

\abstract{%
In this article, we propose the \emph{Artificial Intelligence Security Taxonomy} to systematize the knowledge of threats, vulnerabilities, and security controls of machine-learning-based (ML-based) systems.
We first classify the damage caused by attacks against ML-based systems, define ML-specific security, and discuss its characteristics.
Next, we enumerate all relevant assets and stakeholders and provide a general taxonomy for ML-specific threats.
Then, we collect a wide range of security controls against ML-specific threats
through an extensive review of recent literature. 
Finally, we classify the vulnerabilities and controls of an ML-based system in terms of each vulnerable asset in the system's entire lifecycle.
}


\maketitle

\section{Introduction}
\label{sec:intro}

With the increasing application of artificial intelligence in real life,
the security of machine learning technology 
has been studied actively in recent years, e.g., in 
computer vision~\cite{DBLP:journals/access/AkhtarM18,DBLP:journals/access/AkhtarMKS21},
medical imaging~\cite{DBLP:journals/eswa/KavianiHS22},
speaker recognition~\cite{DBLP:journals/jsa/LanZYWCH22},
malware detection~\cite{DBLP:journals/tissec/DemetrioCBLAR21},
cyber-physical systems~\cite{DBLP:journals/iotj/LiLCXLW20},
and
natural language processing~\cite{DBLP:journals/tist/ZhangSAL20,DBLP:journals/ijon/QiuLZH22}.

Among those academic studies, an enormous amount of literature focuses on specific attack and defense techniques.
One of the best-known security attacks is 
to provide a system with
malicious input data, called \emph{adversarial examples}~\cite{DBLP:journals/corr/SzegedyZSBEGF13}, that cause a malfunction of machine learning models.
When such vulnerable models are used in safety-critical systems, such as autonomous cars and medical devices, adversarial examples may cause fatal accidents.
However, state-of-the-art technologies cannot completely remove the vulnerabilities of models.

Another significant challenge in machine learning security is the vulnerabilities in data collection.
For example, a \emph{data poisoning attack}~\cite{DBLP:journals/tse/HeMCHH22,wang2022threats}
manipulates a training dataset to produce a poisoned model that behaves incorrectly and leads to system malfunction.
To prevent such attacks, we need to protect the system's entire lifecycle, from data collection to system operation.
However, the previous studies have yet to clarify the whole picture of the vulnerabilities in the system lifecycle involving machine learning technologies.

One of the first steps to investigating the landscape of the vulnerabilities in the entire system lifecycle is to build a taxonomy of threats, vulnerabilities, and security controls of ML-based systems (namely, systems using machine learning technologies).
Although various taxonomies on the threats of machine learning have been proposed (e.g.,~\cite{tabassi2019taxonomy,caroline2020artificial,caroline2021securing,mitre:advmlthreatmatrix}),
they do not classify vulnerabilities for each asset in an exhaustive and comprehensive way.
Furthermore, the previous taxonomies 
do not include recent advances in new attack methods, e.g., for resource exhaustion and certain information leakage.

In this paper, we propose the \emph{Artificial Intelligence Security Taxonomy} to systematize the knowledge of threats, vulnerabilities, and security controls of ML-based systems
from the perspectives of information security and software engineering.
Precisely, we classify the damage of ML-specific attacks, present relevant assets and stakeholders, provide a general taxonomy of ML-specific threats, and classify the vulnerabilities and controls of an ML-based system in terms of each vulnerable asset.

\subsection{Contributions}
\label{sub:overview-AISTF}

The main contributions of this paper are as follows:
\begin{itemize}
\item
We propose the \emph{Artificial Intelligence Security Taxonomy} to systematize the knowledge of threats, vulnerabilities, and security controls of ML-based systems
from the perspectives of information 
security and software engineering.
\item
Our taxonomy deals with ML-based systems instead of ML components alone.
Although ``AI security'' and ``ML security'' often refer to the security of ML models, we emphasize that the security of ML technologies should be assessed and controlled across multiple assets in the system lifecycle.
\item 
We follow the conventional approach to information security to make our framework consistent with ISO 27000 series~\cite{isoiec27000:2018}.
We enumerate all relevant assets (\Sec{sub:asset}) and stakeholders (\Sec{sub:stakeholder}) 
and provide a general taxonomy for ML-specific threats (\Sec{sub:T:classification})
that reflects recent advances in new attack methods, 
e.g., sponge attacks aiming at resource exhaustion~\cite{DBLP:conf/eurosp/ShumailovZBPMA21} and poisoning attacks for resource exhaustion~\cite{DBLP:journals/corr/abs-2203-08147} and for information leakage~\cite{DBLP:conf/sp/MahloujifarGC22}.
\item 
We collect a wide range of security controls against ML-specific threats through an extensive review of recent literature.
Then we classify the vulnerabilities and controls of an ML-based system in terms of each vulnerable asset (\Sec{sec:V}).
Based on our classification, we point out areas of potential future research on security control techniques.

\item 
Compared to previous survey papers, our goal is not to explain the mathematical details of attack algorithms.
Instead, we aim to build a general taxonomy 
independent of algorithms and implementations.
\item
Our taxonomy focuses on (centralized) supervised learning.
Its extension to other categories of machine learning
(e.g., unsupervised, semi-supervised, reinforcement, online, or distributed learning)
is left for future work.

\item We have designed the taxonomy to be compatible with our technical report~\cite{MLQM:v2}
that presents the principles and methodologies of the quality management of ML-based systems.
\end{itemize}

\subsection{Related Work}
\label{sub:related}

From a broader perspective, national regulations and guidelines on artificial intelligence have been actively proposed and discussed in recent years.
For example, the European Commission has proposed the Artificial Intelligence Act~\cite{EU:21:AI-act} to regulate the providers of AI systems in a risk-based approach.
The National Security Commission on Artificial Intelligence (NSCAI) in the U.S. Government has provided a report~\cite{NSCAI:21} to show an integrated national strategy in the era of AI-accelerated competition and conflict.
Furthermore, international standards on artificial intelligence technologies have been proposed, e.g., in 
ISO/IEC JTC 1/SC 42~\cite{ISO:22:SC42}.

NIST has been developing AI Risk Management Framework (AI RMF)~\cite{NIST:22:AIRMF}, which shows the characteristics of AI trustworthiness and the principles of the risk management of AI systems.
NIST's draft IR 8269~\cite{tabassi2019taxonomy} provides a taxonomy for AI security in terms of attacks, defenses, and consequences, and a glossary of terminology used in  recent literature on adversarial machine learning.
ENISA's reports on AI cybersecurity~\cite{caroline2020artificial,caroline2021securing}
deal with the threats targeting ML techniques and the vulnerabilities of ML algorithms.

Compared to those prior taxonomies, 
our taxonomy includes a larger class of damage and threats, reflecting recent research advances.
Furthermore, it focuses on ML-specific security and classifies the vulnerabilities and security controls according to each vulnerable asset exhaustively and comprehensively.

Microsoft and MITRE et al. present ATLAS (Adversarial Threat Landscape for Artificial-Intelligence Systems)~\cite{mitre:advmlthreatmatrix} to systematize the threats to ML-based systems and the tactics and techniques at each stage of actual attacks.
Our taxonomy is orthogonal to theirs in that
it provides a general taxonomy of threats, vulnerabilities, and controls that are not specific to the concrete attack tactics shown in their framework.

There are also many papers on the survey and taxonomy of machine learning security (e.g., ~\cite{DBLP:conf/eurosp/PapernotMSW18,
DBLP:journals/access/LiuLZCYL18,
DBLP:journals/jzusc/Li18,
DBLP:journals/jpdc/WangLKTL19,
DBLP:journals/csr/HuangKRSSTWY20,
DBLP:journals/access/XueYWZL20,
9099439,
DBLP:journals/access/LiuXWZXYV21,
kong_survey_2021,
DBLP:journals/sncs/SarkerFN21,
DBLP:journals/tse/HeMCHH22,
MIRSKY2023103006,
DBLP:journals/corr/abs-2207-05164,
DBLP:journals/corr/abs-2201-04736}).
Unlike all previous literature, however, we provide a general taxonomy of ML-specific security in terms of each vulnerable asset.
Furthermore, our taxonomy reflects recent advances in new attack methods (e.g., resource exhaustion and information leakage by data poisoning), which no prior taxonomy has dealt with.

\subsection{Plan of the Paper}
\label{sub:plan}

In \Sec{sec:preliminaries}, we review fundamental concepts from machine learning and information security.
In \Sec{sec:overview}, we provide an overview of ML-specific security in terms of damage to ML-based systems, and define the notion of the ML-specific security used in our taxonomy.
In \Sec{sec:assets:stakeholders}, we show a list of assets and stakeholders in the entire lifecycle of ML-based systems.
In \Sec{sec:attacks:MLBS}, we introduce and explain attacks against ML-based systems.
In particular, we present a classification of ML-specific threats and show the attack surface and possible attackers against ML-based systems.
In \Sec{sec:V}, we provide the list of vulnerabilities and controls for each ML-specific threat.
In \Sec{sec:conclusion}, we summarize the security controls of
ML-based systems for each asset in the systems,
and
present our final remarks.

\section{Preliminaries}
\label{sec:preliminaries}

In this section, we review fundamental concepts from machine learning and information security.

In the rest of this paper, 
we write A/B to represent A or B, possibly both, 
instead of the exclusive disjunction of A and B.
The symbol \Rare{} indicates that the security control has not been studied sufficiently.

\begin{figure}[t]%
\centering
\includegraphics[width=0.97\textwidth]{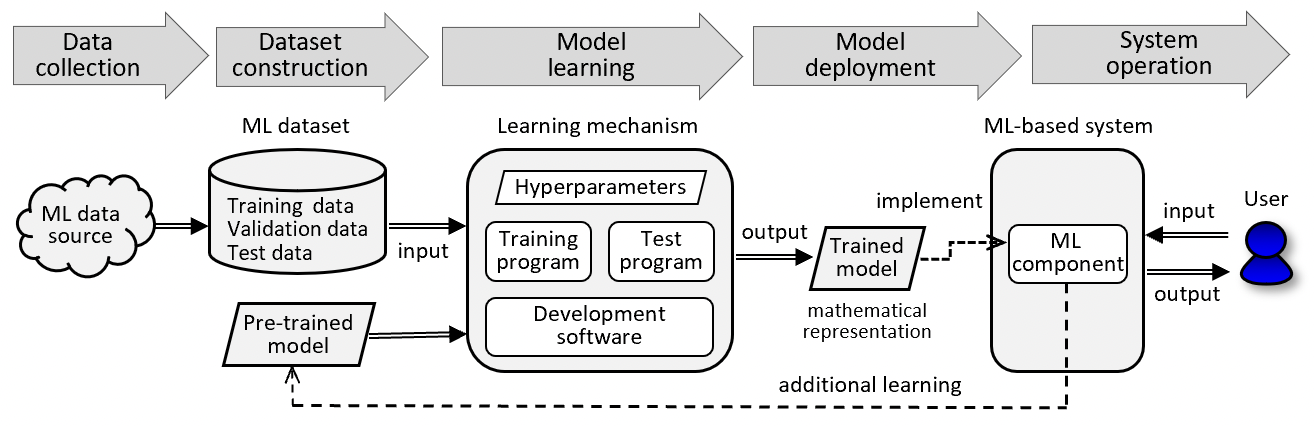}
\caption{Basic components in supervised learning and an ML-based system's lifecycle.}
\label{fig:Fig-ML-basic}
\end{figure}

\subsection{Supervised Machine Learning}
\label{sec:sML}

\emph{Supervised learning} is a category of machine learning that aims to determine a function by using examples of the function's input-output pairs (called \emph{training data})~\cite{book:russell:2010:AI}.
A mathematical representation derived from supervised learning is referred to as a \emph{trained model} (or a \emph{model}).
In \Fig{fig:Fig-ML-basic}, we show the basic components of supervised learning.

For example, \emph{classification}, a type of supervised learning, produces a model that maps a feature vector (input) to a discrete class label (output).
Then this model is used to predict a discrete class label of an \emph{unseen} feature vector.
For another example, \emph{regression} is a type of supervised learning used to predict a continuous value for a given input.
Examples of supervised learning algorithms are decision trees~\cite{song2015decision}, $K$-nearest neighbors~\cite{fix1952discriminatory}, support vector machines~\cite{cortes_support-vector_1995},  Na\"{i}ve Bayes~\cite{hastie2009elements}, logistic regression~\cite{cox1958regression}, and neural networks~\cite{mcculloch1943logical}.

A trained model is specified by information obtained in training (called \emph{parameters}) and information fixed before training (e.g., model architecture).
A \emph{hyperparameter} is a parameter used to control the process of training a model.

\emph{Training data} (resp. \emph{validation/test data}) are correct examples of input-output pairs used to learn (resp. evaluate) a model.
\emph{Validation data} are used to assess a model in tuning hyperparameters during training, whereas \emph{test data} are used to check a model after completing the training.
An \emph{ML dataset} (or a \emph{dataset}) refers to a collection of training, validation, and test data.

\subsection{Machine Learning Based Systems}
\label{sec:MLBS}

A \emph{machine learning component} (for short, an \emph{ML component}) is a software component that implements a trained model.
A \emph{machine-learning-based system} (for short, an \emph{ML-based system} or a \emph{system}) is an information system that executes ML components and uses their output.
A \emph{conventional information system} (or a \emph{non-ML-based system}) is an information system that is not ML-based.
See \Sec{sub:asset} for the details of the structure of an ML-based system.

An ML-based system's lifecycle (\Fig{fig:Fig-ML-basic}) consists of
\emph{system development} and \emph{system operation}.
System development consists of the following phases:
\begin{itemize}
\item \emph{Data collection}: 
Data are collected from ML data sources.
\item \emph{Dataset construction}: 
ML datasets are constructed from collected data.
Typically, this phase includes data cleaning and data pre-processing.
\item \emph{Model learning}: 
Models are designed, trained, validated, tuned, and tested using ML datasets.
Specifically, this phase includes the design and implementation of a model, model training using training data, model validation using validation data, model optimization by hyperparameter tuning, and model evaluation using test data.
\item \emph{Model deployment}: 
ML components are implemented using the trained models.
Then an ML-based system is constructed from the ML components and other software components.
\item \emph{Additional learning}: 
Trained models used in ML components are re-trained, validated, tuned, and tested using ML datasets for additional learning.
\end{itemize}

\subsection{Assets, Threats, Vulnerabilities, and Controls}
\label{sub:basic:security}

\begin{figure}[t]%
\centering
\includegraphics[width=0.80\textwidth]{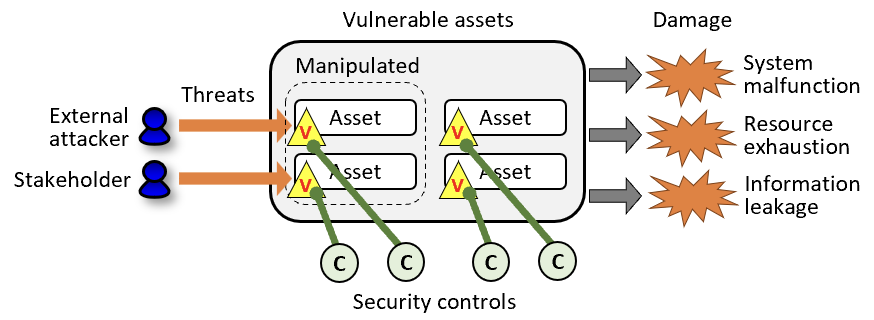}
\caption{Basic concepts in the security of an ML-based system.
V and C denote vulnerabilities and security controls, respectively.}
\label{fig:Fig-IS-basic}
\end{figure}

We review key concepts in information security defined in ISO/IEC 27000 series~\cite{isoiec27000:2018}. 
We show an overview in \Fig{fig:Fig-IS-basic}.

An \emph{asset} is anything that has value to the organization and, therefore, requires protection~\cite{isoiec27005:2018}.
Examples of assets are any data, devices, processes, and systems in the organization.
In the rest of this paper, we deal with assets related to ML-based systems, such as data sources, datasets, trained models, programs, and systems.
See \Sec{sub:asset} for the list of assets in the entire lifecycle of ML-based systems.

A \emph{threat} is a potential cause of the compromise of assets and organizations~\cite{isoiec27005:2018}.
For example, well-known categories of threats are spoofing of user identity, tampering,  repudiation, information disclosure, denial of service, and elevation of privilege~\cite{DBLP:journals/ieeesp/Torr05}.

An asset's \emph{vulnerability} is its weakness that can be exploited by threats to cause damage to the asset, while it does not cause damage in itself~\cite{isoiec27005:2018}.
A \emph{security control} is a measure against a security risk 
where threats will potentially exploit vulnerabilities of assets.

\subsection{Security Risk Assessment}
\label{sub:assess}

For the security of a computing system, performing a  \emph{security risk assessment} for the system's entire lifecycle is essential.

In a security risk assessment, we first identify the stakeholders and assets in the system's lifecycle.
Then we analyze possible threats and vulnerabilities and evaluate their risks.
Since we may not be able to identify all potential threats and vulnerabilities or implement all security controls,
we should prioritize the threats and vulnerabilities and implement security controls in order of priority.
Furthermore, we should perform security risk assessments not only in the design and development of the system, but periodically throughout the system's lifecycle.

In this paper, we do not show a concrete method for a security risk assessment of an ML-based system, but provide a taxonomy of ML-specific security that would help assess an ML-based system.

In a security risk assessment for an ML-based system, we need to identify the threats and vulnerabilities specific to machine learning and those for conventional information systems in the entire system lifecycle.
In the assessment for conventional information systems, the developers should refer to ISO 27000 series~\cite{isoiec27000:2018}, ISO/IEC 15408 (Common Criteria)~\cite{isoiec15408-1:2022}, and NIST SP 800-30~\cite{nistsp800-30}. 
As for control systems in factories and critical infrastructure, the developers should refer to IEC 62443~\cite{iects62443-1-1:2009}.
These security standards have not addressed the ML-specific security this paper deals with.

\section{Overview of ML-Specific Security}
\label{sec:overview}

In this section, we provide an overview of the machine-learning-specific security  (hereafter referred to as \emph{ML-specific security}) of ML-based systems.
Specifically, we explain the characteristics of ML-specific security,
classify the damage caused by attacks on ML-based systems,
and define the notion of the ML-specific security used in our taxonomy.

\subsection{Characteristics of ML-Specific Security}
\label{sub:characteristics}

We first present five motivations (M1 to M5) for our survey and taxonomy by briefly explaining the characteristics of ML-specific security as follows.

\subsubsection{Assessment of threats and vulnerabilities in the system lifecycle}
\label{sub:char1}
We need to identify and evaluate threats that cause damage during system operation by exploiting vulnerabilities in different phases, such as data collection and system development.
For example, data poisoning attacks (\Sec{sub:V:data-poison}) 
manipulate training data during data collection to cause a model's malfunction during its operation.
For another example, backdoor attacks (\Sects{sub:V:data-poison} and \ref{sub:V:model-poison}) require malicious actions both in the system development and in the system operation.

To defend against such attacks across different phases, we need to apply security controls of various assets.
Hence, we figure out the following question:

\shabox{\parbox[l]{0.9\textwidth}{M1: What assets in the system's entire lifecycle should we evaluate to identify threats and vulnerabilities? }}

~\\%
In \Sec{sub:asset}, we present a comprehensive list of assets in the system lifecycle.
We also show how each asset is connected in developing a trained model (\Fig{fig:asset1}) and in operating an ML-based system (\Fig{fig:asset2}).

\subsubsection{Various situations of stakeholders in the system lifecycle}
\label{sub:char2}

In an ML-based system's lifecycle, stakeholders may have opportunities to contaminate or disclose ML data/models and, therefore, can be potential attackers against the system.
However, the system lifecycle tends to involve various stakeholders in different situations.
For instance, ML datasets, pre-trained models, and learning mechanisms, respectively, may or not be provided by other third parties.
Therefore, these assets may be attacked by different attackers in various situations.

To identify and evaluate possible attackers in various situations, 
we work on the following question:

\shabox{\parbox[l]{0.9\textwidth}{ M2: What stakeholders and their situations should we consider to figure out possible attackers in the system's entire lifecycle? }}

~\\%
In \Sec{sub:stakeholder}, we present a list of stakeholders (\Tbl{tab:principals}) and a list of their situations in the system lifecycle (\Tbl{tab:situations}) to clarify what stakeholders provide each asset and may manipulate it.
Then, based on \Tbl{tab:situations}, we derive the list of possible attackers for each threat (\Tbl{tab:threats:AS}) in \Sec{sub:attack-surface}.

\subsubsection{Security controls at the system level in the presence of vulnerable models}
\label{sub:char3}
We cannot completely remove a trained model's vulnerabilities by applying a state-of-the-art method of model development.
For example, despite many studies on adversarial examples
~\cite{biggio2018wild,DBLP:journals/caaitrit/ChakrabortyADCM21,ijcai2021p635,electronics11081283,DBLP:journals/air/AldahdoohHFD22,DBLP:journals/access/KhamaisehBAMA22},
no machine learning algorithm is known to produce models that behave correctly for all adversarial examples (\Sec{sub:V:evasion}).

To mitigate the damage caused by attacks, 
we can 
apply security controls also on assets other than the trained model in the system.
For example, an access control program helps restrict the input to a trained model, hence reducing the risk of the trained model's vulnerabilities (\Sec{sub:V:all-oracles}).

To cope with vulnerable models at the system level,
we figure out the following question:

\shabox{\parbox[l]{0.9\textwidth}{ M3: What security controls at the system level can we apply to an ML-based system? }}

~\\%
In \Sec{sub:V:all-oracles}, we present the system-level vulnerabilities 
against the input of malicious data to systems during operation.
In \Sec{sub:V:all-threats}, we show the system-level vulnerabilities against ML-specific threats in general.

\subsubsection{Multiple layers of security controls to hidden threats}
\label{sub:char4}
We often cannot detect the threats to a trained model used in a system.
For example, some poisoning attacks embed backdoors into a trained model that cannot be detected easily (\Sec{sub:V:model-poison}).
Thus, the system operators may not notice such threats until the system is damaged by the threats.

To counter such hidden threats, we need to design multiple layers of security controls.
For example, when developers cannot trust datasets or pre-trained models, they may need to pre-process them even if detection techniques cannot identify poisoning attacks  (\Sects{sub:V:data-poison} and~\ref{sub:V:model-poison}).

In general, a single security control tends to be insufficient to achieve the security of ML-based systems.
Therefore, we need to figure out the security controls to different assets.

\shabox{\parbox[l]{0.9\textwidth}{ M4: What security control techniques can we choose to design multiple layers of controls for each ML-specific threat? }}

~\\%
To address this question, in \Sec{sec:conclusion}, we summarize the security controls against ML-specific threats for each asset (\Tbls{tab:control:1}, \ref{tab:control:2}, \ref{tab:control:3}, and \ref{tab:control:op}).
Enumerating possible choices of security controls against ML-specific threats would be a good starting point for designing multiple layers of controls.

\subsubsection{Missing security controls}
\label{sub:char5}

As we have discussed, ML-based systems may require various security controls for various assets.
Considering many combinations of assets, stakeholder situations, threats, and vulnerabilities, 
AI security research so far may not have covered all means of security controls.

\shabox{\parbox[l]{0.9\textwidth}{ M5: What kinds of security controls are missing in AI security research? }}

~\\%
To address this question, we classify the vulnerabilities and security controls in terms of each vulnerable asset.
Then we find missing pieces of security controls that have not been studied sufficiently and may be an area of potential future research.
We mark them with the symbol \Rare{} in each table on security controls.

\subsection{Damage to ML-Based Systems}
\label{sub:damage}

We classify the damage caused by attacks on ML-based systems as follows:
\begin{enumerate}
\item Loss of integrity/availability (\Sec{sub:damage:IA})
\begin{enumerate}
\item System malfunction due to the unintended behavior of an ML component
\item System malfunction due to other factors
\item Exhaustion of resources by an ML component
\item Exhaustion of resources by conventional software/hardware
\end{enumerate}
\item Loss of confidentiality (\Sec{sub:damage:C})
\begin{enumerate}
\item[(e)] Leakage of information on a trained model
\item[(f)] Leakage of sensitive information in a training dataset
\item[(g)] Leakage of other information
\end{enumerate}
\end{enumerate}
In Table~\ref{tab:damage-attacks}, we summarize threats that cause these damages.

\begin{table}[t]
  \caption{Damage caused by attacks against ML-based systems.}
  \label{tab:damage-attacks}
\begin{footnotesize}
\begin{tabular}{@{}l@{}l@{}l@{}l@{}}\hlineb \\[-1.5ex]
\textgt{Damage} & & \textgt{Threats} & \textgt{} \\\cline{3-4}
 & & \textgt{ML-specific threats} & \textgt{Other threats} \\
 & & & (out of scope) \\[0.5ex]
\hline \\[-1.5ex]
 \multirow{4}{*}{\begin{tabular}{@{}l@{}} Loss of \\ integrity/ \\ availability \\ (\Sec{sub:damage:IA}) \end{tabular}}
 & System malfunction 
\\[0.5ex]
 & \multirow{3}{*}{\begin{tabular}{@{}l@{}} (a) due to the unintended ~\\\,~~~ behavior of an ML \\\,~~~ component \end{tabular}} & \multirow{2}{*}{\begin{tabular}{@{}l@{}} Data poisoning attack \\  (\Sec{sub:V:data-poison}) \end{tabular}} & \multirow{5}{*}{\begin{tabular}{@{}l@{}} Conventional threat \\ against the software/ \\ hardware used to \\ implement the ML \\ component \end{tabular}} \\
 & & & \\
 & & \multirow{2}{*}{\begin{tabular}{@{}l@{}} Model poisoning attack \\  (\Sec{sub:V:model-poison}) \end{tabular}} & \\
 & & &  \\
 & & Evasion attack (\Sec{sub:V:evasion}) &  \\
\cline{3-4}
 & \multirow{1}{*}{\begin{tabular}{@{}l@{}} (b) due to other factors \end{tabular}} & & \multirow{2}{*}{\begin{tabular}{@{}l@{}} Conventional threat \\ against the system \end{tabular}} \\
 & & & \\
\cline{2-4}
 & \multirow{1}{*}{\begin{tabular}{@{}l@{}} Resource exhaustion \end{tabular}} & \\[0.5ex]
 & (c) by an ML component ~& \multirow{2}{*}{\begin{tabular}{@{}l@{}} Data poisoning attack \\  (resource exhaustion; \Sec{sub:V:data-poison}) \end{tabular}} ~& \multirow{5}{*}{\begin{tabular}{@{}l@{}} Conventional threat \\ against the software/ \\ hardware used to \\ implement the ML \\ component \end{tabular}} \\
 & & & \\
 & & \multirow{2}{*}{\begin{tabular}{@{}l@{}} Model poisoning attack \\  (resource exhaustion; \Sec{sub:V:model-poison}) \end{tabular}} & \\
 & & & \\
 & & Sponge attack (\Sec{sub:V:sponge}) & \\
\cline{3-4}
 & (d) \multirow{2}{*}{\begin{tabular}{@{}l@{}} by conventional \\ software/hardware \end{tabular}} & & \multirow{2}{*}{\begin{tabular}{@{}l@{}} Conventional threat \\ against the system \end{tabular}} \\
\\[0.5ex]
\hline \\[-1.5ex]
 \multirow{3}{*}{\begin{tabular}{@{}l@{}} Loss of \\ confidentiality \\ (\Sec{sub:damage:C}) ~\end{tabular}}
 &  \multirow{1}{*}{\begin{tabular}{@{}l@{}} Leakage of \end{tabular}} &  & \\[0.5ex]
 &  \multirow{2}{*}{\begin{tabular}{@{}l@{}} (e) information on a~ \\~~~\, trained model \end{tabular}} & Model extraction attack & \multirow{2}{*}{\begin{tabular}{@{}l@{}} Conventional threat \\ of model thefts \end{tabular}} \\
 & & (\Sec{sub:V:model-extract}) & \\
 \cline{3-4}
 & \multirow{2}{*}{\begin{tabular}{@{}l@{}} (f) sensitive information \\~~~\, in a training dataset \end{tabular}} & 
 \multirow{2}{*}{\begin{tabular}{@{}l@{}} Information leakage attack \\ of training data (\Sec{sub:V:leakage-data}) \end{tabular}} & \multirow{2}{*}{\begin{tabular}{@{}l@{}} Conventional threat \\ of data thefts \end{tabular}} \\
 & & & \\
 & & Data poisoning attack & \\
 & & (information embedding; \Sec{sub:V:data-poison}) &  
\\[0.5ex]
\cline{3-4}
 & \multirow{2}{*}{\begin{tabular}{@{}l@{}} (g) other confidential \\~~~\,\, information \end{tabular}} & \multirow{2}{*}{\begin{tabular}{@{}l@{}} Model poisoning attack \\ (information embedding; \Sec{sub:V:model-poison}) \end{tabular}} ~& \multirow{2}{*}{\begin{tabular}{@{}l@{}} Conventional threat \\ of data thefts \end{tabular}} \\
 & & & \\[0.5ex]
 \hlineb
\end{tabular}
\end{footnotesize}
\end{table}

\subsubsection{Loss of Integrity/Availability}
\label{sub:damage:IA}

Attacks against an ML-based system may result in the loss of integrity/availability by malfunctioning the system.
Such a system malfunction is caused by
either (a) the unintended behavior of an ML component
or (b) other factors owing to conventional threats.
For example, an autonomous car's accidents may be caused by a malfunction of an ML component for object detection, and also by a malfunction of conventional software.

The unintended behavior of an ML component may be caused by 
(a1) the malfunction of the trained model, including the malfunction of the model's interpretation functionality, 
(a2) an unintended functionality of the trained model, 
or (a3) a conventional (non-ML-specific) threat against the software/hardware used to implement the ML component (e.g., buffer overrun and fault injection attacks).

In this paper, we focus on factors (a1) and (a2), namely, ML-specific threats, such as data poisoning attacks (\Sec{sub:V:data-poison}), model poisoning attacks (\Sec{sub:V:model-poison}), and evasion attacks (\Sec{sub:V:evasion}).
We do not address the details of (a3) conventional (non-ML-specific) threats,
because we can apply security controls for (a3) in an analogous way to conventional information systems.

Attacks against an ML-based system may also result in the loss of availability due to the exhaustion of resources by either (c) an ML component or (d) conventional software/hardware.
The former (c) may be caused by 
(c1) a trained model of the ML component
or (c2) a conventional threat against the software/hardware used to implement the ML component.
Again, we focus only on (c1) ML-specific threats, called sponge attacks (\Sec{sub:V:sponge}).

\subsubsection{Loss of Confidentiality}
\label{sub:damage:C}

Attacks against an ML-based system may result in the loss of confidentiality owing to information leakage from trained models.
For example, if an ML component is trained using a sensitive dataset about patients, then a medical information system using the ML component may breach patient privacy owing to the information leakage of training data via access to the ML component.

There are 
three
types of ML-specific threats to confidentiality: 
\begin{itemize}
\item[] (e) those to extract information on a trained model (model extraction attacks in \Sec{sub:V:model-extract});
\item[] (f) those to obtain sensitive information in the training dataset from the behavior of the trained model (information leakage attacks of training data in \Sec{sub:V:leakage-data});
\item[] (g) those to embed sensitive information into trained models to disclose it during system operation
(information embedding attacks in \Sec{sub:V:model-poison}).
\end{itemize}
In these threats, examples of sensitive information include personal information, trade secrets, and information that violates laws, regulations, or contracts.
Remarkably, (f) may involve 
data poisoning attacks in some cases (information embedding attacks in \Sec{sub:V:data-poison}).

Again, we focus only on ML-specific threats to confidentiality in this paper.
We do not address the details of conventional threats of model/data thefts,
which may exploit vulnerabilities in the development software, the development environment, the conventional software components in the system, the computing environment, and the operating organization.
As a measure to prevent conventional data theft, \emph{secure multi-party computation} is a technology for computing encrypted data without decrypting them, and has been actively studied for its applications to machine learning~\cite{DBLP:conf/sp/MohasselZ17,DBLP:conf/ndss/PatraS20,DBLP:journals/popets/AttrapadungHIKM22}.

\subsection{ML-Specific Security}
\label{sub:ML-spec:security}

We define the notion of \emph{ML-specific security} (or the scope of this paper) as
the security of ML-based systems against the risks 
arising \emph{via
trained models} that result in unintended behavior, resource exhaustion, or information leakage.

This notion does \emph{not} refer to the entire security of ML-based systems (often called ``security for AI''), but deals with the security against the attacks by which a trained model behaves unintendedly, exhausts resources, or leaks sensitive information.

Our definition of ML-specific security corresponds to the ML-specific threats shown in \Tbl{tab:damage-attacks}.
For example, ML-specific security deals with the risks associated with the following damage to ML-based systems:
\begin{itemize}
\item[](a') System malfunction due to a \emph{trained model}'s unintended behavior;
\item[](c') Resource exhaustion by a \emph{trained model};
\item[](e') Leakage of information on a trained model through  
the \emph{trained model}'s inputs and outputs;
\item[](f') Leakage of sensitive information in a training dataset through 
a \emph{trained model}'s inputs and outputs.
\end{itemize}

Since the \emph{non-ML-specific security} of information systems has been widely studied and standardized (e.g., as ISO/IEC 27000~\cite{isoiec27000:2018}),
our survey and taxonomy focus only on ML-specific security and do not cover the non-ML-specific security of conventional software used in systems.

For example, the ML-security does \emph{not} deal with risks associated with the following damage:
\begin{itemize}
\item[](a'') System malfunction caused only by the conventional software or hardware used to implement ML components;
\item[](c'') Resource exhaustion caused only by the conventional software or hardware used to implement ML components;
\item[](e'') Trained model theft by unauthorized access to the system or the computing environment during system operation;
\item[](f'') Training data theft by unauthorized access to the development environment.
\end{itemize}

Finally, we remark that 
the security achieved by ML technologies (often called ``security by AI'')
is a different notion from the ML-specific security in this paper,
whereas ML technologies (e.g., detection of malicious input to ML-based systems) are occasionally used to achieve the ML-specific security of systems.

\section{Assets and Stakeholders in the Lifecycle of ML-Based Systems}
\label{sec:assets:stakeholders}

In this section, we show a list of assets, stakeholders, and their situations in the lifecycle of an ML-based system.

\begin{figure}[t]%
\centering
\includegraphics[width=0.80\textwidth]{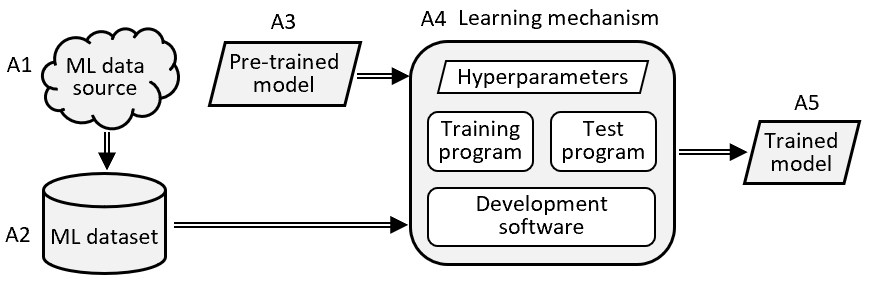}
\caption{Assets in the development of a trained model.}
\label{fig:asset1}
\end{figure}

\begin{figure}[t]%
\centering
\includegraphics[width=0.99\textwidth]{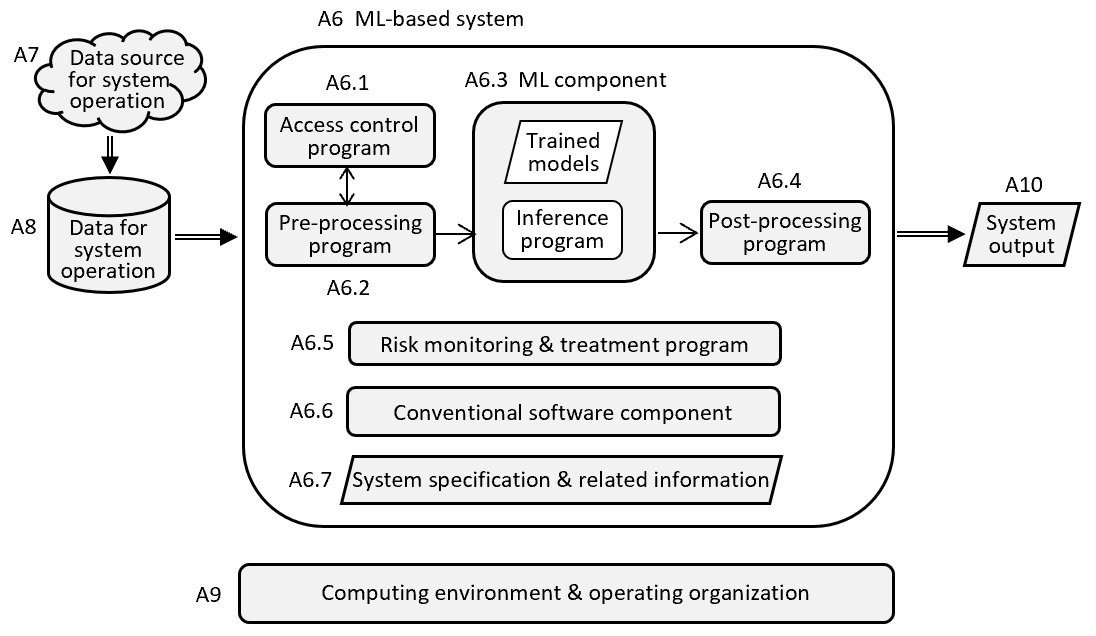}
\caption{Assets in the operation of an ML-based system.}
\label{fig:asset2}
\end{figure}

\begin{table}[t]
  \caption{List of assets in development.}
  \label{tab:asset:develop}
\begin{footnotesize}%
\begin{tabular}{@{}l@{}l@{}l@{}l@{}l@{}l@{}}\hlineb
\multicolumn{2}{@{}l}{\textgt{Asset}} & \textgt{Description} \\\hline
A1 ~~& ML data source & \multirow{2}{*}{\begin{tabular}{@{}l@{}} A population, a process, or an environment from which raw data instances \\ are collected to construct an ML dataset. \end{tabular}} 
\\
\\
\hline
A2 & ML dataset & \multirow{2}{*}{\begin{tabular}{@{}l@{}} A collection of data instances that is obtained by pre-processing raw data \\ and is used to train and test a model. \end{tabular}} 
\\
\\
\hline
A3 & Pre-trained model & \multirow{2}{*}{\begin{tabular}{@{}l@{}} A trained model that third parties have developed in advance and provided \\ for other developers. \end{tabular}} 
\\
\\
\hline
A4 & Learning mechanism ~& \multirow{4}{*}{\begin{tabular}{@{}l@{}} A software component for developing a model from an ML dataset  (and \\ possibly a pre-trained model) by using machine learning technologies; \\
typically consisting of hyperparameters, training programs, test programs, \\ and development software. \end{tabular}} 
\\
\\
\\
\\
\hline
A5 & \multirow{2}{*}{\begin{tabular}{@{}l@{}} Trained model \\ (or model) \end{tabular}} & \multirow{2}{*}{\begin{tabular}{@{}l@{}} A mathematical representation produced by a learning mechanism using \\ an ML dataset. \end{tabular}} 
\\
\\
\hline
A6 & ML-based system & \multirow{1}{*}{\begin{tabular}{@{}l@{}} An information system that executes ML components and uses their output. \end{tabular}}
\\
\hline
A11 & \multirow{2}{*}{\begin{tabular}{@{}l@{}} ML data source for \\ additional learning \end{tabular}} & \multirow{2}{*}{\begin{tabular}{@{}l@{}} A data source used to construct a dataset for additional learning after the \\ system's operation. \end{tabular}}
\\
\\
\hline
A12 & \multirow{2}{*}{\begin{tabular}{@{}l@{}} ML dataset for \\ additional learning \end{tabular}} & \multirow{2}{*}{\begin{tabular}{@{}l@{}} A dataset used for additional learning after the system's operation. \end{tabular}}
\\
\\
\hlineb
\end{tabular}
\end{footnotesize}
\end{table}

\begin{table}[t]
  \caption{List of assets in an ML-based system.}
  \label{tab:asset:ML-component}
\begin{footnotesize}%
\begin{tabular}{@{}l@{}l@{}l@{}l@{}l@{}l@{}}\hlineb
\multicolumn{2}{@{}l}{\textgt{Asset}} & \textgt{Description} \\\hline
A6.1 ~~& \multirow{2}{*}{\begin{tabular}{@{}l@{}} Access control \\ program \end{tabular}} & A program that controls the input of data for system operation. \\
\\
\hline
A6.2 & \multirow{2}{*}{\begin{tabular}{@{}l@{}} Pre-processing \\ program \end{tabular}} & \multirow{2}{*}{\begin{tabular}{@{}l@{}} A program that processes raw data to produce input to ML components. \\
(This may access the ML components' internal information or may be \\\,
combined with the ML components.) \end{tabular}} \\
\\
\\
\hline
A6.3 & ML component & \multirow{2}{*}{\begin{tabular}{@{}l@{}} A software component that implements a trained model and possibly \\
its interpretation functionality. \end{tabular}} \\
\\
\hline
A6.4 & \multirow{2}{*}{\begin{tabular}{@{}l@{}} Post-processing \\ program \end{tabular}} & \multirow{2}{*}{\begin{tabular}{@{}l@{}} A program that processes the ML component's output and its \\ interpretation. \end{tabular}} \\
\\
\hline
A6.5 & \multirow{2}{*}{\begin{tabular}{@{}l@{}} Monitoring/risk \\ treatment program \end{tabular}} & A program that treats risks by monitoring the system's behavior. \\
\\
\hline
A6.6 & \multirow{2}{*}{\begin{tabular}{@{}l@{}} Other conventional ~\\ software components \end{tabular}} & Other software components that do not consist of ML components. \\
\\
\hline
A6.7 & \multirow{2}{*}{\begin{tabular}{@{}l@{}} System specification \\ \& related information \end{tabular}} ~& \multirow{2}{*}{\begin{tabular}{@{}l@{}} Information on the ML datasets, the trained models, the other system \\ specifications, and their related information, such as datasets or models \\ resembling the ones used in the system development or operation. \end{tabular}}
\\
\\
\\
\hlineb
\end{tabular}
\end{footnotesize}
\end{table}

\begin{table}[t]
  \caption{List of assets in operation.}
  \label{tab:asset:operation}
\begin{footnotesize}%
\begin{tabular}{@{}l@{}l@{}l@{}l@{}l@{}l@{}}\hlineb
\multicolumn{2}{@{}l}{\textgt{Asset}} & \textgt{Description} \\\hline
A7 ~~~& \multirow{2}{*}{\begin{tabular}{@{}l@{}} Data source for \\ system operation \end{tabular}} & \multirow{2}{*}{\begin{tabular}{@{}l@{}} A population, a process, or an environment from which raw data \\ instances are collected and used as input to ML-based systems. \end{tabular}}
\\
\\
\hline
A8 & \multirow{2}{*}{\begin{tabular}{@{}l@{}} Data for system \\ operation \end{tabular}} & \multirow{2}{*}{\begin{tabular}{@{}l@{}} A set of data instances input to ML-based systems during operation. \end{tabular}}
\\
\\
\hline
A9 & \multirow{3}{*}{\begin{tabular}{@{}l@{}} Computing environment \\ \& operating organization ~\\ during system operation \end{tabular}} & \multirow{2}{*}{\begin{tabular}{@{}l@{}} The computing environment used by the ML-based system and \\ the organization that operates the system. \end{tabular}}
\\
\\
\\
\hline
A10 & System output & Data that an ML-based system outputs.
\\
\hlineb
\end{tabular}
\end{footnotesize}
\end{table}

\subsection{Assets in ML-Based Systems}
\label{sub:asset}

Identifying assets in the system lifecycle is essential to evaluate their threats and vulnerabilities 
(\Sec{sub:char1}).
We show the assets in the development of a trained model 
in \Fig{fig:asset1} 
and 
those in the operation of an ML-based system
in \Fig{fig:asset2}.
These figures also indicate the data flow across assets to reason about how a manipulated asset has an impact on another asset.

In \Tbls{tab:asset:develop}, \ref{tab:asset:ML-component}, and \ref{tab:asset:operation},
we briefly explain the assets in system development (A1 to A6 and A11 to A12), those in an ML-based system (A6.$*$), and those in system operation (A7 to A10), respectively.

\subsection{Stakeholders and Their Situations in ML-Based Systems}
\label{sub:stakeholder}

\begin{table}[t]
  \caption{List of stakeholders and description of their actions.}
  \label{tab:principals}
\begin{footnotesize}%
\begin{tabular}{@{}l@{}l@{}l@{}l@{}l@{}l@{}}\hlineb
\multicolumn{2}{@{}l}{\textgt{Stakeholder}} & \textgt{Description} \\\hline
P1 ~~& Manager of ML data sources & Actor who manages ML data sources \\
\hline
P2 & ML Data provider & Actor who provides ML datasets \\
\hline
P3 & Model provider & Actor who provides pre-trained models \\
\hline
P4 & Learning mechanism provider ~ & Actor who provides learning mechanisms \\
\hline
P5 & System developer & Actor who develops ML-based systems \\
\hline
P6 & \multirow{2}{*}{\begin{tabular}{@{}l@{}} Manager of data sources for \\[-0.3ex] system operation \end{tabular}} & Actor who manages data sources for system operation \\
\\
\hline
P7 & \multirow{2}{*}{\begin{tabular}{@{}l@{}} Data provider for system \\[-0.3ex] operation \end{tabular}} & Actor who provides data for system operation \\
\\
\hline
P8 & System operator & \multirow{2}{*}{\begin{tabular}{@{}l@{}} Actor who operates ML-based systems to use them or  \\[-0.3ex] to provide their services for system end-users \end{tabular}} \\
\\
\hline
P9 & System end-user & \multirow{2}{*}{\begin{tabular}{@{}l@{}} Actor who provides ML-based systems with input data \\[-0.3ex] to use their services \end{tabular}} \\
\\
\hline
P10 & \multirow{2}{*}{\begin{tabular}{@{}l@{}} Manager of data sources for \\[-0.3ex] additional learning \end{tabular}} & \multirow{2}{*}{\begin{tabular}{@{}l@{}} Actor who manages data sources for the additional  \\[-0.3ex] learning of models used in the ML-based systems \end{tabular}} \\
\\
\hline
P11 & \multirow{2}{*}{\begin{tabular}{@{}l@{}} Data provider for additional \\[-0.3ex] learning \end{tabular}} & \multirow{2}{*}{\begin{tabular}{@{}l@{}} Actor who provides data for the additional learning of \\[-0.3ex] models used in the ML-based systems \end{tabular}}
\\
\\
\hline
P12 & Model user & \multirow{2}{*}{\begin{tabular}{@{}l@{}} Actor who is provided pre-trained models and uses \\[-0.3ex] the models to develop new ones \end{tabular}} 
\\
\\
\hlineb
\end{tabular}
\end{footnotesize}
\end{table}

Next, we present a list of stakeholders and their situations.
Identifying stakeholders and understanding their situations is essential to evaluate possible attackers in the system's entire lifecycle 
(\Sec{sub:char2}).

\subsubsection{Actions by Stakeholders}
\label{sub:stakeholder:def}

We show the list of possible stakeholders (P1 to P12) in ML-based systems in \Tbl{tab:principals}.
Some stakeholders may not exist in a specific system's lifecycle.

\emph{Managers of data sources} (P1, P6, P10) manage the data sources; e.g., researchers manage the scientific experiments from which they obtain data.
However, they do not exist when nobody is able to manage the data sources in external environments;
e.g., no stakeholder may be able to control the situation on roadsides from which autonomous cars obtain data.

\emph{Data providers} (P2, P7, P11) provide data for development or operation, and can be data owners, data curators, data engineers, or data brokers.
\emph{Learning mechanism providers} (P4) include providers of training programs, test programs, and development software.
\emph{System developers} (P5) develop ML-based systems either by using models pre-trained by model providers or by developing training models from ML datasets (and pre-trained models in the case of transfer learning).
They may also conduct additional learning to update a model if the model's quality has decreased after operation.

\emph{System operators} (P8) either use ML-based systems on their own or provide services to system end-users.
\emph{System end-users} (P9) query input into ML-based systems to use the systems' services.
They may not exist when system operators use the systems by themselves.

\emph{Model users} (P12) are third parties for whom developers provide pre-trained models.
They use the provided models to develop new models.

\begin{table}[t]
  \caption{Situations of stakeholders that provide each asset in the system lifecycle.
  }
  \label{tab:situations}
\begin{footnotesize}%
\spB%
\begin{tabular}{@{}l@{}l@{}l@{}l@{}l@{}l@{}l@{}}\hlineb
\multicolumn{2}{@{}l}{\textgt{Asset}} & \multicolumn{4}{@{}l}{\textgt{Stakeholders that provide the asset}} & \spB\textgt{Situation} \\
 &  & \multicolumn{2}{@{}l}{System development phase} ~&  \multicolumn{2}{@{}l}{System operation phase} & \\
\hline
A1 ~~~& ML data source & P1 ~~& \begin{tabular}{@{}l@{}} Manager of ML \\ data sources \end{tabular} & ~~~~~~&  & S1a \\
\cline{3-7}
 &  &  & (Nobody) &  & & S1b \\
\hline
A2 & ML dataset & P2 & ML data provider &  &  & S2a \\
\cline{3-7}
 &   & P5 & System developer &  & & S2b \\
\hline
A3 & Pre-trained model & P3 & Model provider &  & & S3a  \\
\cline{3-7}
 &  &  & (Nobody) &  &  & S3b\\
\hline
A4 & Learning mechanism  & P4 & \begin{tabular}{@{}l@{}} Learning mechanism \\ provider \end{tabular} &  & & S4a \\
\cline{3-7}
 &   & P5 & System developer &  & & S4b \\
\hline
A5 & Trained model & P5 & System developer &  &  & S5 \\
\hline
A6 & System  & P5 & System developer & P8 & System operator & S6 \\
\hline
A7 & \multirow{2}{*}{\begin{tabular}{@{}l@{}} Data source for \\ system operation \end{tabular}}  &  &  & P6 & \begin{tabular}{@{}l@{}} Manager of data sources ~\\ for system operation \end{tabular} ~& S7a \\
\cline{3-7}
 &  &  &  &  & (Nobody) & S7b \\
\hline
A8 & \multirow{2}{*}{\begin{tabular}{@{}l@{}} Data for system \\ operation \end{tabular}}  &  &  & P7 & \begin{tabular}{@{}l@{}} Data provider for \\ system operation \end{tabular} & S8a \\
\cline{3-7}
 &   &  &  & P9 & System end-user & S8b \\
\cline{3-7}
 &   &  &  & P8 & System operator & S8c \\
\hline
A9 & \multirow{2}{*}{\begin{tabular}{@{}l@{}} Computing environment \\ \& operating organization  \end{tabular}} ~~ &  &  & P8 & System operator & S9 \\
\\
\hline
A10 & System's output data  &  &  &  & (Nobody) & S10 \\
\hline
A11 & \multirow{2}{*}{\begin{tabular}{@{}l@{}} ML data source \\ for additional learning \end{tabular}}  & P10 ~~& \begin{tabular}{@{}l@{}} Manager of ML data \\ sources for additional \\ learning \end{tabular} &  & & S11a \\
\cline{3-7}
 &  &  & (Nobody) &  & & S11b \\
\hline
A12 & \multirow{2}{*}{\begin{tabular}{@{}l@{}} ML dataset for \\ additional learning \end{tabular}}  & P11 & \begin{tabular}{@{}l@{}} Data provider for \\ additional learning \end{tabular} &  & & S12a \\
\cline{3-7}
 &  & P5 & System developer &  & & S12b \\
\cline{3-7}
 &  &  & (Nobody) &  &  & S12c \\
\hlineb
\end{tabular}
\end{footnotesize}
\end{table}

\subsubsection{Situations of Stakeholders}
\label{sub:stakeholder:sit}
We next present the situations of stakeholders from the viewpoint of the assets in the system lifecycle.
In \Tbl{tab:situations}, we show which stakeholders provide each asset during system development and during system operation.
For each $i = 1, 2, \ldots , 12$, a symbol of the form S$i*$ denotes a situation where the asset A$i$ is provided by a certain stakeholder.
For example, 
S2a denotes the situation where A2 (an ML dataset) is provided by P2 (ML data providers),
while S2b denotes the situation where A2 is provided by P5 (the system developer).

We can use each row of \Tbl{tab:situations} to identify which stakeholder can manipulate each asset as an attacker.
For instance, in situations S2a and S2b, P2 (the ML data provider) and P5 (the system developer), respectively, may be an attacker that manipulates A2 (the ML dataset).

\Tbl{tab:situations} shows that there are hundreds\footnote{A simple calculation shows the number of combinations of situations is around 500 ($\approx 1^4 \times 2^6 \times 3^2$), since four assets have single situations, six have two situations, and two have three situations in  \Tbl{tab:situations}.} of combinations of the situations of asset provision in the entire system lifecycle that we deal with.
Therefore, there are a large number of potential attack scenarios corresponding to these stakeholder situations.
Based on \Tbl{tab:situations}, we will derive the list of possible attackers for each threat (\Tbl{tab:threats:AS}) in \Sec{sub:attack-surface}.

\section{Attacks Against ML-Based Systems}
\label{sec:attacks:MLBS}

In this section, we introduce and explain attacks against ML-based systems.
We first present a classification of threats to ML-based systems.
Then we show the attack surface and possible attackers against an ML-based system.
Finally, we remark on the attacker's knowledge of an ML-based system under attack,
which determines the attack's feasibility and efficiency.

\subsection{Taxonomy of ML-Specific Threats}
\label{sub:T:classification}
In our taxonomy, we focus on \emph{ML-specific} 
and \emph{deliberate} threats.
Namely, we deal with neither conventional (non-ML-specific) threats 
nor threats that arise from negligence, accidents, or environmental factors.
To deal with conventional threats and non-deliberate threats, we should apply conventional security controls analogously to traditional information security.

Our taxonomy of ML-specific threats focuses on three typical situations: \emph{development}, \emph{operation}, and \emph{model provision}, as follows.
\begin{itemize}
\item[] (T1) ML-specific threats to ML-based systems during development;
\item[] (T2) ML-specific threats to ML-based systems during operation;
\item[] (T3) ML-specific threats to pre-trained models provided for model users.
\end{itemize}

(T1) and (T2), respectively, deal with threats before and during system operation. 
(T3) deals with pre-trained models alone and not with an ML-based system.

We remark that this paper focuses on ML-specific threats and does not deal with the details of conventional security threats to ML-based systems
(e.g., the theft of ML datasets provided for developers are conventional threats, as shown in~\Tbl{tab:damage-attacks}).

\begin{table}[t]
  \caption{Classification of threats to ML-based systems during development,
  where C, I, and A denote the loss of confidentiality, integrity, and availability, respectively.}
  \label{tab:threats-ML-SD}
\begin{footnotesize}%
\spB%
\begin{tabular}{@{}l@{}l@{}l@{}c@{}l@{}}\hlineb
\multicolumn{2}{@{}l@{}}{\textgt{Threat}} & \textgt{Sub-threat} & \spB\textgt{Damage} ~& \textgt{Description} \\\hline
\\[-1.8ex]
T1.1 ~& \multirow{3}{*}{\begin{tabular}{@{}l@{}} Data \\[-0.3ex] poisoning ~\\[-0.3ex] attack \end{tabular}} &  &  & Manipulation of an ML data source or an ML dataset  \\
& & Malfunction & I, A & -- to cause a malfunction of the trained model \\
& & ~~Targeted & &  ~~~$*$ for specific inputs \\
& & ~~Backdoor & &  ~~~$*$ for inputs that contain specific information \\
& & ~~Non-targeted~~ & &  ~~~$*$ for unspecified inputs \\[0.7ex]
& & \multirow{2}{*}{\begin{tabular}{@{}l@{}} Functionality \\[-0.3ex] change \end{tabular}} & I, A & -- to obtain a trained model with an unintended functionality \\[-0.3ex]
& & &  &  \\[0.7ex]
& & \multirow{2}{*}{\begin{tabular}{@{}l@{}} Resource \\[-0.3ex] exhaustion \end{tabular}} & A~~~ & -- to cause the exhaustion of resources during system operation \\
\\
& & \multirow{2}{*}{\begin{tabular}{@{}l@{}} Information \\[-0.3ex] embedding \end{tabular}} & C~~~ & \multirow{2}{*}{\begin{tabular}{@{}l@{}} -- to embed sensitive information into ML datasets to disclose \\ ~\, it during system operation \end{tabular}} \\[-0.3ex]
& & &  &  \\
\hline
T1.2 & \multirow{3}{*}{\begin{tabular}{@{}l@{}} Model \\[-0.3ex] poisoning \\[-0.3ex] attack \end{tabular}} &  &  & Manipulation of a pre-trained model, a learning mechanism,  \\[-0.3ex]
& & & & or a trained model \\
& & Malfunction & I, A & -- to cause a malfunction of a trained model \\
& & ~~Targeted & & ~~~$*$ for specific inputs \\
& & ~~Backdoor & & ~~~$*$ for inputs that contain specific information \\
& & ~~Non-targeted~~ & & ~~~$*$ for unspecified inputs \\[0.7ex]
& & \multirow{2}{*}{\begin{tabular}{@{}l@{}} Functionality \\[-0.3ex] change \end{tabular}} & I, A & -- to obtain a trained model with an unintended functionality \\[-0.3ex]
& & &  &  \\[0.7ex]
& & \multirow{2}{*}{\begin{tabular}{@{}l@{}} Resource \\[-0.3ex] exhaustion \end{tabular}} & A ~~~& -- to cause the exhaustion of resources during system operation \\
\\
& & \multirow{2}{*}{\begin{tabular}{@{}l@{}} Information \\[-0.3ex] embedding \end{tabular}}  & C ~~~& -- to embed sensitive information into model parameters or \\[-0.3ex]
& & & & ~\, hyperparameters to disclose them during system operation  \\\hlineb
\end{tabular}
\end{footnotesize}
\end{table}

\begin{table}[t]
  \caption{Classification of threats to ML-based systems during operation,
  where C, I, and A denote the loss of confidentiality, integrity, and availability, respectively.}
  \label{tab:threats-ML-SO}
\begin{footnotesize}%
\spB%
\begin{tabular}{@{}l@{}l@{}l@{}c@{}l@{}c@{}}\hlineb \\[-1.8ex]
\multicolumn{2}{@{}l}{\textgt{Threat}} & \textgt{Sub-threat} & \spB\textgt{Damage} ~& \textgt{Description} \\\hline
\\[-1.8ex]
T2.1 ~& \multirow{3}{*}{\begin{tabular}{@{}l} Exploitation \\[-0.3ex] of a poisoned \\[-0.3ex] model  \\[-0.3ex]~ \end{tabular}} &  &  & \begin{tabular}{@{}l} Input of malicious data to a trained model to \\[-0.3ex] exploit poisoning to cause \end{tabular}  \\
& & For system malfunction & I, A & -- the model's unintended behavior \\
& & For resource exhaustion & A ~~~& -- the exhaustion of resources by the model \\
& & For information leakage & C~~~ & \multirow{1}{*}{\begin{tabular}{@{}l} -- the leakage of sensitive information from the model \end{tabular}} \\
\hline
T2.2 & \multirow{2}{*}{\begin{tabular}{@{}l@{}} Model \\[-0.3ex] extraction \\[-0.3ex] attack \end{tabular}} &  &  & \begin{tabular}{@{}l} Input of malicious data to a trained model to \\ cause the leakage of \end{tabular} \\
&  & For model attributes & C~~~ & -- attributes of the trained model \\[-0.3ex]
&  & For model functionalities & C~~~ & -- functionalities of the trained model \\
\hline
T2.3 & \multirow{2}{*}{\begin{tabular}{@{}l@{}} Evasion \\[-0.3ex] attack \end{tabular}} &  &  &  \multirow{2}{*}{\begin{tabular}{@{}l} Input of adversarial examples to a trained model \\[-0.3ex] to cause a malfunction of the trained model \end{tabular}} \\
\\
&  & Targeted & I, A & -- \multirow{1}{*}{\begin{tabular}{@{}l@{}} for specific inputs during operation \end{tabular}} \\
&  & Indiscriminate & I, A & -- \multirow{1}{*}{\begin{tabular}{@{}l@{}} for unspecified inputs during operation \end{tabular}} \\
\hline
T2.4 & \multirow{2}{*}{\begin{tabular}{@{}l@{}} Sponge \\[-0.3ex] attack \end{tabular}} &   & A ~~~& \multirow{2}{*}{\begin{tabular}{@{}l@{}} Input of sponge examples to a trained model \\[-0.3ex] to cause resource exhaustion during operation \end{tabular}} \\
\\
\hline
T2.5 & \multirow{4}{*}{\begin{tabular}{@{}l@{}} Information \\[-0.3ex] leakage attack\spB\!\\[-0.3ex] of training  \\[-0.3ex] data \end{tabular}} &  &  & \multirow{3}{*}{\begin{tabular}{@{}l@{}} Input of malicious data to a trained model \\[-0.3ex] to cause the leakage of sensitive information \\[-0.3ex] in a training dataset used to train the model \end{tabular}} \\
& & Membership inference & C~~~  \\[-0.3ex]
&  & Attribute inference & C~~~ \\[-0.3ex]
&  & Data reconstruction & C~~~ \\[-0.3ex]
&  & Property inference & C~~~ \\\hlineb
\end{tabular}
\end{footnotesize}
\end{table}

\subsubsection{ML-Specific Threats in System Development}
ML-specific threats in system development are poisoning attacks against the assets used in the development.
In \Tbl{tab:threats-ML-SD}, we show the classification of (T1).

We define ML-specific threats during development as follows:
\begin{itemize}
\item A \emph{data poisoning attack} (\Sec{sub:V:data-poison}) is an attack that manipulates an ML data source or an ML dataset 
to cause the trained model's unintended behavior, the exhaustion of resources by the trained model, or the leakage of sensitive information from the trained model.
\item A \emph{model poisoning attack} (\Sec{sub:V:model-poison}) is an attack that manipulates a pre-trained model, a learning mechanism, or a trained model
to cause the trained model's unintended behavior, the exhaustion of resources by the trained model, or the leakage of sensitive information from the trained model.
\end{itemize}
Compared to the previous taxonomy papers, our definition clarifies that poisoning attacks may cause resource exhaustion and information leakage.

\subsubsection{ML-Specific Threats in System Operation}
\label{sub:T:classification:op}

ML-specific threats in system operation are malicious inputs to systems during operation.
In \Tbl{tab:threats-ML-SO}, we show the classification of (T2).

We define ML-specific threats during operation as follows:
\begin{itemize}
\item An \emph{exploitation of a poisoned model} (\Sec{sub:V:EPM}) is an attack that inputs malicious data during operation to exploit poisoning 
to cause the trained model's unintended behavior, the exhaustion of resources by the trained model, or the leakage of sensitive information from the trained model.
\item A \emph{model extraction attack} (\Sec{sub:V:model-extract}) is an attack that inputs malicious data to a trained model during operation to cause the leakage of information on the trained model.
\item An \emph{evasion attack} (\Sec{sub:V:evasion}) is an attack that inputs adversarial examples to a trained model to cause a malfunction of the trained model.
\item A \emph{sponge attack} (\Sec{sub:V:sponge}) is an attack that inputs malicious data to a trained model to cause the exhaustion of resources during system operation.
\item An \emph{information leakage attack of training data} (\Sec{sub:V:leakage-data}) is an attack that inputs malicious data to a trained model during operation to cause the leakage of sensitive information in a training dataset used to train the model.
\end{itemize}

\subsubsection{ML-Specific Threats in Model Provision}
ML-specific threats to a pre-trained model provided for a model user deal with a situation where
an attacker against the pre-trained model is 
either the model provider or the model user (who is not the model developer).
In this paper, the threats in model provision are:
\begin{itemize}
\item (T1.2) model poisoning attacks (\Sec{sub:V:model-poison}) by the model provider who aims to cause the malfunction of a model developed by the model user; 
\item (T3) white-box \emph{information leakage attacks of training data}  (\Sec{sub:V:leakage-data}) by the model user who aims to obtain information in the training dataset (used by the model provider) from the pre-trained model.
\end{itemize}

We remark that a black-box information leakage attack of training data is the threat (T2.5) in system operation (\Sec{sub:T:classification:op}).

\subsection{Attack Surface and Attackers}
\label{sub:attack-surface}

We show the attack surface of an ML-based system and the possible attackers against the system.
Although this paper focuses on ML-specific security, attacks against ML-based systems may combine both ML-specific attacks and conventional security attacks.
In particular, we explain that the attack surface for an ML-specific threat is often exploited by a pre-attack against a vulnerability in conventional software or information system.

\begin{table}[t]
  \caption{Assets on the attack surface and possible attackers in ML-specific threats, where Ext, Dev, and Op, respectively, denote external attackers, system developers, and system operators. 
  Environmental factors are omitted from this table.}
  \label{tab:threats:AS}
\begin{footnotesize}
\begin{tabular}{@{}l@{}l@{}l@{}l@{}l@{}c@{}c@{}c@{}l@{}l@{}}\hlineb
\multirow{2}{*}{\begin{tabular}{@{}l@{}} \textgt{Attack} \\[-0.5ex] \textgt{situation}\end{tabular}} & \multicolumn{2}{@{}l}{\multirow{2}{*}{\begin{tabular}{@{}l@{}} \textgt{ML-specific} \\ \textgt{Threat}\end{tabular}}} & \multicolumn{2}{@{}l}{\multirow{2}{*}{\begin{tabular}{@{}l@{}} \textgt{Assets on the} \\[-0.5ex] \textgt{attack surface}\end{tabular}}} & \multicolumn{4}{l}{\spB\textgt{Attackers}~~~} \\[-0.2ex]
 &  &  &  &  & Ext.~ & Dev.~ & Op.~ & \multicolumn{2}{@{}l}{Others} \\[0.2ex]
\hline
\multirow{1}{*}{\begin{tabular}{@{}l@{}} Development \end{tabular}} ~& T1.1 ~& \multirow{2}{*}{\begin{tabular}{@{}l@{}} Data \\[-0.3ex] poisoning\end{tabular}} ~& \begin{tabular}{@{}l@{}} A1, \\ A11 \end{tabular} ~& \begin{tabular}{@{}l@{}} ML data source (for \\[-0.3ex] additional learning) \end{tabular} & \checkmark & & & \begin{tabular}{@{}l@{}} P1, \\ P10 \end{tabular} ~& \begin{tabular}{@{}l@{}} Manager of ML data sources \\[-0.3ex] (for additional learning) \end{tabular} \\[-0.2ex]
\cline{4-10}
 & & & \begin{tabular}{@{}l@{}} A2, \\ A12 \end{tabular} ~& \begin{tabular}{@{}l@{}} ML dataset (for \\[-0.3ex] additional learning) \end{tabular} & \checkmark & \checkmark & & \begin{tabular}{@{}l@{}} P2, \\ P11 \end{tabular} ~& \begin{tabular}{@{}l@{}} ML data provider (for  \\[-0.3ex] additional learning) \end{tabular} \\[-0.2ex]
\cline{2-10}
 & T1.2 & \multirow{2}{*}{\begin{tabular}{@{}l@{}} Model \\[-0.3ex] poisoning\end{tabular}} & A3 & Pre-trained model & \checkmark & \checkmark & & P3 & Model provider \\[-0.2ex]
\cline{4-10}
&  &  & A4 & \begin{tabular}{@{}l@{}} Learning \\ mechanism \end{tabular} & \checkmark & \checkmark & & P4 & \begin{tabular}{@{}l@{}} Learning mechanism \\[-0.3ex] provider \end{tabular} \\[-0.2ex]
\cline{4-10}
 &  &  & A5 & Trained model & \checkmark & \checkmark & &  \\
\hline
\multirow{1}{*}{\begin{tabular}{@{}l@{}} Operation\end{tabular}} & T2 &  \multirow{2}{*}{\begin{tabular}{@{}l@{}} Malicious \\[-0.3ex] input of \\[-0.3ex] data for \\[-0.3ex] system \\[-0.3ex] operation \end{tabular}}
 & A6 & System & \checkmark & \checkmark  & \checkmark & P9 & System end-user \\[-0.2ex]
\cline{4-10}
 &  &  & A7 & \begin{tabular}{@{}l@{}} Data source for \\[-0.3ex] system operation\end{tabular} & \checkmark & & & P6 & \begin{tabular}{@{}l@{}} Manager of data sources \\[-0.3ex] for system operation\end{tabular} \\[-0.2ex]
\cline{4-10}
 &  &  & A8 & \multirow{2}{*}{\begin{tabular}{@{}l@{}} Data for system \\[-0.3ex] operation \end{tabular}} & \multirow{2}{*}{\begin{tabular}{@{}l@{}} \checkmark \end{tabular}} &  & \multirow{2}{*}{\begin{tabular}{@{}l@{}} \checkmark \end{tabular}} & P7 & \multirow{2}{*}{\begin{tabular}{@{}l@{}} Data provider for \\[-0.3ex] system operation  \end{tabular}} \\[-0.3ex]
 \\[-0.3ex]
\cline{9-10}
 &  &  &  &  &  &  &  & P9 & System end-user \\[0.0ex]
\hline
\begin{tabular}{@{}l@{}} Model \\[-0.3ex] provision \end{tabular} & T1.2 & \begin{tabular}{@{}l@{}} Model \\[-0.3ex] poisoning\end{tabular} & A3 & Pre-trained model & \checkmark & \checkmark & & \\[-0.2ex]
\cline{2-10}
& \begin{tabular}{@{}l@{}} T3 \end{tabular} & \begin{tabular}{@{}l@{}} Malicious \\[-0.3ex] data input \\[-0.3ex] to ML \\[-0.3ex] component \end{tabular} & A3 & Pre-trained model & \checkmark & & & P12 & Model user \\
\hlineb
\end{tabular}
\end{footnotesize}
\end{table}

\subsubsection{Manipulated Assets and Possible Attackers}
\label{sub:asset-attacker}

A system's \emph{attack surface}~\cite{iects62443-1-1:2009} is its physical and functional interfaces that can be accessed and potentially exploited by an attacker.
In \Tbl{tab:threats:AS}, we show the assets on the attack surface and possible attackers for each threat to an ML-based system.

The possibility of each stakeholder being an attacker (shown in \Tbl{tab:threats:AS}) is derived from \Tbl{tab:situations} in \Sec{sub:stakeholder:sit}, 
which shows the stakeholders that provide each asset and, therefore, can be potential attackers.
For example, P2 (the ML data provider) can be an attacker against A2 (ML datasets), as situation S2a in \Tbl{tab:situations} shows that P2 provides A2 and may manipulate it.

In system development, the attack surface consists of the assets involved in the process of model learning.
Data poisoning attacks (\Sec{sub:V:data-poison}) exploit A1, A11 (ML data sources) or A2, A12 (ML datasets);
model poisoning attacks (\Sec{sub:V:model-poison}) manipulate A3 (pre-trained models), A4 (learning mechanisms), or A5 (trained models).

In system operation, the attack surface consists of the assets involved in the input to the system.
An attacker may input malicious data to A6 (the system) or manipulate A7 (data source for system operation) or A8 (data for system operation)
in the exploitation of poisoned models (\Sec{sub:V:EPM}), model extraction attacks (\Sec{sub:V:model-extract}), evasion attacks (\Sec{sub:V:evasion}), sponge attacks (\Sec{sub:V:sponge}),  and information leakage attacks of training data (\Sec{sub:V:leakage-data}).

Stakeholders during development can be attackers when they provide A1, A11 (ML data sources), A2, A12 (ML datasets), A3 (pre-trained models), or A4 (learning mechanisms) to the developer.
End-users can be attackers during system operation when they can query malicious input to the system.

Internal attackers in ML-specific threats are also shown in \Tbl{tab:threats:AS}.
System developers can be attackers against A2, A12 (ML datasets), A3 (pre-trained models), A4 (learning mechanisms), A5 (trained models), and A6 (systems).
System operators can be attackers against A6 (systems) and A8 (data for system operation).

Environmental factors are omitted from this table, since they should be handled in an analogous way as the case of conventional information systems.

\subsubsection{Pre-Attacks for ML-Specific Attacks}\label{sub:pre-attack}

To exploit assets of an ML-based system, an attacker may need to conduct pre-attacks against conventional software components or information systems.

In system development, 
manipulation of assets in data/model poisoning 
can be conducted via pre-attacks that exploit vulnerabilities in, e.g., data sources, data provision frameworks, development software, or development environments.

For instance, model poisoning may be mounted by exploiting a vulnerability in a software library for machine learning (e.g., TensorFlow or PyTorch) to install backdoors in the library or in the trained model.

In system operation, evasion attacks and information leakage attacks may manipulate (sources of) data for system operation through pre-attacks that exploit vulnerabilities in the system, the computing environment, or the operating organization during system operation.

For instance, an attacker may execute (white-box) evasion attacks or information leakage attacks of training data after stealing trained models, e.g., by unauthorized access to the system, reverse engineering of the trained model, or side-channel attacks.

\subsection{Information Available to Attackers}\label{sub:attacker}
The feasibility and efficiency of ML-specific attacks often depend on the system's information available to attackers.
We provide an overview of an attacker's 
knowledge gained from access
to ML components during operation (\Sec{sub:info:access}) and an attacker's prior knowledge of models, datasets, and specification information that can be used for ML-specific attacks (\Sec{sub:info:models:datasets}).

\subsubsection{Accessibility to ML Components}\label{sub:info:access}
ML-specific attacks are classified in terms of the attacker's knowledge:
\begin{itemize}
\item A \emph{white-box attack} is an attack that uses the knowledge of the architecture and parameters of a trained model;
\item A \emph{black-box attack} is an attack that uses no knowledge of the architecture and parameters of a trained model;
\begin{itemize}
\item An \emph{interactive black-box attack} is a black-box attack where an attacker can query multiple inputs to a trained model;
\item A \emph{blind black-box attack} is a black-box attack where an attacker cannot query multiple inputs to a trained model;
\end{itemize}
\item A \emph{gray-box attack} is between white-box and black-box attacks.
\end{itemize}

A white-box attack assumes a situation where an attacker has access to a trained model's architecture and parameters, for example: 
\begin{itemize}
\item[] (1) the trained model is publicly available;
\item[] (2) the attacker has stolen the trained model's parameters in advance.
\end{itemize}
As for case (1), to prevent or mitigate the white-box attacks, the system developers should pre-process or fine-tune the publicly available trained model before deploying it into the system.
In case (2), to steal parameters and other information of the trained model, attackers need to conduct pre-attacks that exploit vulnerabilities in the development software, the development environment, the system, the computing environment, and the operating organization.
Therefore, to prevent those pre-attacks, the system developer and operator should apply conventional security controls to the assets and the organization.

An interactive black-box attack assumes a situation where an attacker does not know the parameters of the trained model but can input data into the trained model.
Typically, the attacker provides input to the model or the system under attack via legitimate access to inference APIs, and observes the input-output relation to learn the functionality of the trained model.

A blind black-box attack assumes a situation where an attacker neither knows the parameters of the trained model nor queries input to the model.
Although attackers cannot learn the functionality of the trained model, they may generate malicious data (e.g., adversarial examples) in advance using other trained models and inputs them into the system under attack.
These malicious inputs lead to successful attacks when they have transferability across different models.

We remark that the definitions of these attacks can be slightly different depending on their threat types and attack scenarios.
For example, interactive black-box attackers in sponge attacks (\Sec{sub:V:sponge}) can measure the time and energy consumption of the ML component or the system during operation.

\subsubsection{Prior Knowledge of Models, Datasets, and Specification}
\label{sub:info:models:datasets}
Information on trained models or on resembling ones can be used to produce malicious inputs to ML components or ML-based systems during operation.
For instance, it can be used to generate adversarial examples~\cite{DBLP:journals/corr/SzegedyZSBEGF13} for evasion attacks (\Sec{sub:V:evasion}), and to generate inputs to the models in information leakage attacks (\Sec{sub:V:leakage-data}). 
Thus, preventing attackers from obtaining information on the trained models or resembling models may be helpful in mitigating those attacks.

The training dataset or resembling one can be used to build a model that approximates the trained model under attack, and can be used to mount ML-specific attacks, such as evasion attacks.
Therefore, preventing attackers from obtaining the training dataset or resembling one may help mitigate those attacks.
It should be noted that evasion attacks can be more successful if the model is trained using only a single publicly available dataset.

Finally, we remark that specification information on learning mechanisms can also be used to mount ML-specific attacks.
For example, data poisoning attacks (\Sec{sub:V:data-poison}) may exploit information about the learning algorithm and hyperparameters used to train the model.

\section{ML-Specific Threats, Vulnerabilities, and Controls}
\label{sec:V}

In this section, we explain the ML-specific threats and show the vulnerabilities and security controls for each threat.

As mentioned in \Sec{sub:char3}, applying security controls from a system-level perspective is essential to achieve the security of an ML-based system in the presence of vulnerable models.
Furthermore, as mentioned in \Sec{sub:char4}, designing multiple layers of security controls is effective in treating the risks caused by hidden threats.

To identify the vulnerabilities and possible security controls exhaustively and systematically, we classify them in terms of the assets vulnerable to each threat.
It should be noted that vulnerable assets are not limited to the ones that attackers can manipulate directly.
By tracking back the data flow between assets, we enumerate vulnerable assets.
Then we point out the security controls that have not been studied sufficiently (marked with \Rare{} in tables), which are either hard to be implemented or have the potential for future research.

As mentioned in \Sec{sub:assess}, the developers and operators may not be able to implement all security controls.
Therefore, they should prioritize the threats and vulnerabilities of a specific system
and implement security controls in order of priority.

\begin{table}[t]
  \caption{List of tables about the vulnerabilities and controls to ML-specific threats.
  The symbol \Rare{} indicates that security controls have not been studied sufficiently.}
  \label{tab:VC:overview}
\begin{footnotesize}
\begin{tabular}{@{}l@{}l@{}l@{}l@{}l@{}l@{}l@{}}\hlineb
\textgt{Sect.~} & \textgt{Table~~~~} & \multicolumn{2}{@{}l}{\textgt{Threat}} & \textgt{Vulnerable assets}~ & \textgt{Vulnerabilities} & \textgt{Controls} \\\hline
\ref{sub:V:data-poison} & \Tbl{tab:VC:data-poison} & T1.1 ~& \multirow{2}{*}{\begin{tabular}{@{}l} Data poisoning \\ attacks \end{tabular}} & \multirow{2}{*}{\begin{tabular}{@{}l} A1, A2, A4, A5, \\ A11, A12 \end{tabular}} & \multirow{2}{*}{\begin{tabular}{@{}l} V1.1, V1.2, V1.3, \\ V2.1, V2.2, V2.3, \\ V2.4, V4.1, V5.3 \end{tabular}} & \multirow{3}{*}{\begin{tabular}{@{}l} C1.1, C1.2, C1.3, \\ C2.1, C2.2, C2.3, \\ C2.4a, C4.1, C5.3 \end{tabular}} \\
\\
\\
\hline
\ref{sub:V:model-poison} & \Tbl{tab:VC:model-poison} & T1.2 & \multirow{2}{*}{\begin{tabular}{@{}l} Model poisoning \\ attacks \end{tabular}} & A3, A4, A5 & \multirow{4}{*}{\begin{tabular}{@{}l} V3.1, V3.2, V3.3, \\ V3.4, V4.2, V4.3, \\ V4.4, V5.1, V5.2, \\ V5.3 \end{tabular}} & \multirow{4}{*}{\begin{tabular}{@{}l} C3.1, C3.2, C3.3, \\ C3.4, C4.2, C4.3, \\ C4.4, C5.1, C5.2, \\ C5.3 \end{tabular}} \\
\\
\\
\\
\hline
\ref{sub:V:EPM} & \Tbl{tab:VC:EPM} & T2.1 & \multirow{2}{*}{\begin{tabular}{@{}l} Exploitation of \\ poisoned models \end{tabular}} & A6.2, A6.3, A6.4 & V6.2, V6.3a, V6.4 & \multirow{2}{*}{\begin{tabular}{@{}l} C6.2, C1-C5, C6.4\,\Rare{} \\~ \end{tabular}} \\
\\
\hline
\ref{sub:V:model-extract} & \Tbl{tab:VC:model-extract} & T2.2 & \multirow{2}{*}{\begin{tabular}{@{}l} Model extraction \\ attacks \end{tabular}} & A6.2, A6.3, A6.4 & V6.2, V6.3b, V6.4 & \multirow{2}{*}{\begin{tabular}{@{}l} C6.2, C2.4b\,\Rare{}, C4.5b\,\Rare{}, \\ C5.4b, C5.5b\,\Rare{}, C6.4 \end{tabular}} \\
\\
\hline
\ref{sub:V:evasion} & \Tbl{tab:VC:evasion} & T2.3 & Evasion attacks & A6.2, A6.3, A6.4 & V6.2, V6.3c, V6.4 & \multirow{2}{*}{\begin{tabular}{@{}l} C6.2, C2.4c, C4.5c, \\ C5.4c, C5.5c, C6.4 \end{tabular}} \\
\\
\hline
\ref{sub:V:sponge} & \Tbl{tab:VC:sponge} & T2.4 & Sponge attacks & A6.2, A6.3, A6.4 & V6.2, V6.3d, V6.4 & \multirow{3}{*}{\begin{tabular}{@{}l} C6.2\,\Rare{}, C2.4d\,\Rare{}, \\ C4.5d\,\Rare{}, C5.4d\,\Rare{}, \\ C5.5d\,\Rare{}, C6.4\,\Rare{} \end{tabular}} \\
\\
\\
\hline
\ref{sub:V:leakage-data} & \Tbl{tab:VC:leakage-data} & \multirow{2}{*}{\begin{tabular}{@{}l} T2.5 \\ T3 \end{tabular}} & \multirow{3}{*}{\begin{tabular}{@{}l} Information \\ leakage attacks \\ of training data \end{tabular}} & A6.2, A6.3, A6.4 & V6.2, V6.3e, V6.4 & \multirow{2}{*}{\begin{tabular}{@{}l} C6.2\,\Rare{}, C2.4e, C4.5e, \\ C5.4e, C5.5e, C6.4 \end{tabular}} \\
\\
\\
\hline
\ref{sub:V:all-oracles} & \Tbl{tab:VC:all-MIS} & T2.$*$ & \multirow{3}{*}{\begin{tabular}{@{}l} All malicious \\ input during \\ system operation \end{tabular}} & A6.1, A7, A8 & \multirow{3}{*}{\begin{tabular}{@{}l} V6.1, V7.1, V7.2, \\ V7.3, V8.1, V8.2, \\ V8.3 \end{tabular}} & \multirow{3}{*}{\begin{tabular}{@{}l} C6.1, C7.1, C7.2, \\ C7.3, C8.1, C8.2, \\ C8.3 \end{tabular}} \\
\\
\\
\hline
\ref{sub:V:all-threats} & \Tbl{tab:VC:all-threats} & T$*$ & \multirow{2}{*}{\begin{tabular}{@{}l} All ML-specific \\ threats \end{tabular}} & \multirow{2}{*}{\begin{tabular}{@{}l} A6.5, A6.6, \\ A6.7, A9 \end{tabular}} & \multirow{2}{*}{\begin{tabular}{@{}l} V6.5, V6.6, V6.7, \\ V9.1, V9.2, V9.3 \end{tabular}} & \multirow{2}{*}{\begin{tabular}{@{}l} C6.5, C6.6, C6.7, \\ C9.1, C9.2, C9.3 \end{tabular}} \\
\\
\hlineb
\end{tabular}
\end{footnotesize}
\end{table}

\subsection{Outline}
\label{sub:V:overview}

\begin{itemize}
\item
First, we explain the ML-specific threats \emph{during system development}
and show the vulnerabilities and security controls for these threats.
Specifically, we deal with data poisoning attacks (\Sec{sub:V:data-poison}) and model poisoning attacks (\Sec{sub:V:model-poison}).

\item
Second, we explain the ML-specific threats \emph{during system operation}
and show the vulnerabilities and security controls for these threats.
Specifically, we deal with the exploitation of poisoned models (\Sec{sub:V:EPM}), model extraction attacks (\Sec{sub:V:model-extract}), evasion attacks (\Sec{sub:V:evasion}), sponge attacks (\Sec{sub:V:sponge}), and information leakage attacks of training data (\Sec{sub:V:leakage-data}). 

\item
Finally, we show the system-level vulnerabilities and controls common in different types of threats
(\Sects{sub:V:all-oracles} and \ref{sub:V:all-threats}).
\end{itemize}

In \Tbl{tab:VC:overview}, we list the tables about the vulnerable assets, vulnerabilities, and controls to ML-specific threats.

\subsection{Data Poisoning Attacks}
\label{sub:V:data-poison}

\subsubsection{Overview}
A \emph{data poisoning attack} is an attack that manipulates an ML data source or an ML dataset 
to cause the trained model's unintended behavior, the exhaustion of resources by the trained model, or the leakage of sensitive information from the trained model.

\begin{figure}[t]%
\centering
\includegraphics[width=0.98\textwidth]{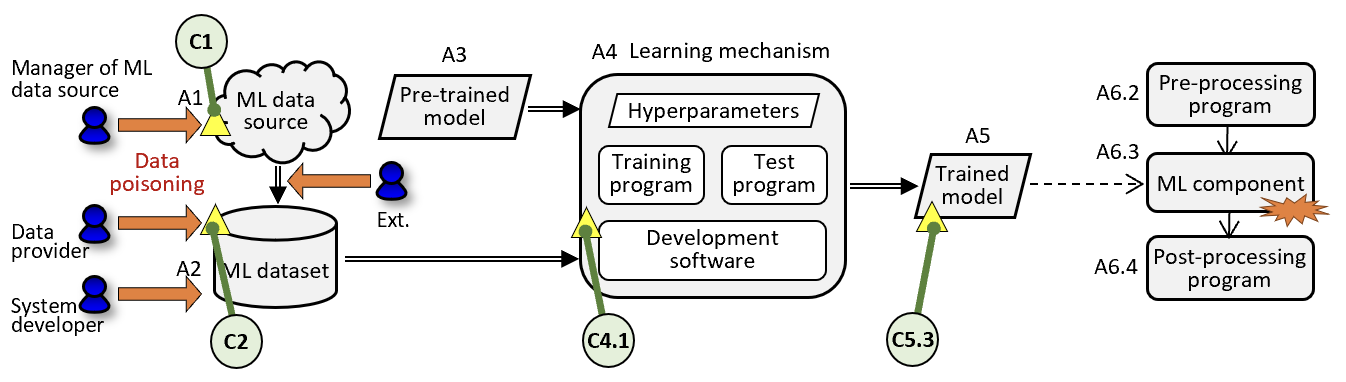}
\caption{Vulnerable assets and security controls against  data poisoning attacks.
 Ext denotes an external attacker.}
\label{fig:data-poison}
\end{figure}

\begin{table}[t]
  \caption{Vulnerabilities and controls to \emph{data poisoning attacks} (T2.1).
  See \Tbl{tab:VC:all-threats} for vulnerabilities and controls common to all attack types.}
  \label{tab:VC:data-poison}
\begin{footnotesize}
\begin{tabular}{@{}l@{}l@{}l@{}l@{}l@{}l@{}}\hlineb
\multicolumn{2}{l}{\spB\textgt{Vulnerable asset}} & \multicolumn{2}{l}{\spB\textgt{Vulnerability}} & \multicolumn{2}{l}{\spB\textgt{Control}} \\\hline
\spB\begin{tabular}{l} A1, \\ A11 \end{tabular} & \begin{tabular}{l} ML data \\ source \end{tabular} & V1.1 & \begin{tabular}{l} Lack of a process to \\ evaluate the trustworthiness \\ of ML data sources \end{tabular} & C1.1 & \begin{tabular}{l} Evaluate the trustworthiness \\ of ML data sources \end{tabular} \\
\cline{3-6}
 &  & V1.2 & \begin{tabular}{l} Lack of security controls \\ to prevent/mitigate the \\ poisoning of ML data \\ sources \end{tabular} & C1.2 & \begin{tabular}{l} Apply security controls to \\ prevent/mitigate the poisoning \\ of ML data sources \end{tabular} \\
\cline{3-6}
 &  & V1.3 & \begin{tabular}{l} Lack of a process to detect \\ the poisoning of ML data \\ sources \end{tabular} & C1.3 & \begin{tabular}{l} Use techniques to detect the \\  poisoning of ML data sources \end{tabular} \\
\hline
\spB\begin{tabular}{l} A2, \\ A12 \end{tabular} & \begin{tabular}{l} ML \\ dataset \end{tabular} & V2.1 & \begin{tabular}{l} Lack of a process to evaluate \\ the trustworthiness of ML \\ datasets \end{tabular} & C2.1 & \begin{tabular}{l} Evaluate the trustworthiness \\ of ML datasets \end{tabular} \\
\cline{3-6}
 &  & V2.2 & \begin{tabular}{l} Lack of security controls \\ to prevent/mitigate the \\ poisoning of ML datasets \end{tabular} & C2.2 & \begin{tabular}{l} Apply security controls to \\ prevent/mitigate the poisoning \\ of ML datasets \end{tabular} \\
\cline{3-6}
 &  & V2.3 & \begin{tabular}{l} Lack of a process to detect \\ the poisoning of ML datasets \end{tabular} & C2.3 & \begin{tabular}{l} Use techniques to detect the \\ poisoning of ML datasets \end{tabular} \\
\cline{3-6}
 &  & V2.4 & \begin{tabular}{l} Lack of a process to make \\ ML datasets resilient to \\ data poisoning \end{tabular} & C2.4a & \begin{tabular}{l} Use techniques to synthesize/\\ pre-process the ML datasets \\ to make them resilient to data \\ poisoning \end{tabular} \\
\hline
A4 & \begin{tabular}{l} Learning \\ mechanism \end{tabular} & V4.1 & \begin{tabular}{l} Learning mechanisms being \\ vulnerable to data poisoning \end{tabular} & C4.1 & \begin{tabular}{l} Use learning mechanisms being \\ more resilient to data poisoning \end{tabular} \\
\hline
A5 & \begin{tabular}{l} Trained \\ model \end{tabular} & V5.3 & \begin{tabular}{l} Lack of a process to remove/ \\ reduce poisoning effects from \\ trained models \end{tabular} & C5.3 & \begin{tabular}{l} Use techniques to remove/\\ reduce poisoning effects from \\ trained models \end{tabular} \\
\hlineb
\end{tabular}
\end{footnotesize}
\end{table}

\subparagraph{Damage}
This kind of attack is conducted during data collection, dataset construction, or model learning. 
The damage caused by this attack occurs during system operation.
A data poisoning attack can malfunction the system or exhaust resources, and decreases the system's qualities, such as performance, safety, and fairness~\cite{DBLP:conf/pkdd/SolansB020,DBLP:conf/aaai/MehrabiNMG21}. 
It can also induce the leakage of sensitive information~\cite{DBLP:conf/sp/MahloujifarGC22} and may also cause privacy breaches.

\subparagraph{Attack situation}
We show an overview of the attack situations in data poisoning attacks in \Fig{fig:data-poison}.

Manipulations of training datasets in data poisoning attacks are classified into (i) data injection, (ii) data modification, and (iii) label manipulation~\cite{wang2022threats}.
Data poisoning attacks may manipulate not only training data, but also validation/test data so that the developer can miss the attacks.

Typical attackers and their situations are as follows:
\begin{itemize}
\item A third party (e.g., an external attacker or a data curator) manipulates the ML data sources (A1);
\item The data provider manipulates the ML dataset (A2);
\item A third party (e.g., an external attacker) manipulates the ML dataset (A2) in dataset construction or model learning.
\end{itemize}

\subparagraph{Classification w.r.t. damage}
We list the following major types of data poisoning attacks:
\begin{enumerate}
\item[(a)] \emph{Malfunction attacks}
\begin{itemize}
\item \emph{Targeted attacks}: Data poisoning attacks that cause a malfunction of a trained model for specific inputs to the model;
\item \emph{Backdoor attacks}: Data poisoning attacks that cause a malfunction of a trained model for inputs that contain specific information (trigger);
\item \emph{Non-targeted attacks}: Data poisoning attacks that cause a malfunction of a trained model for unspecified inputs;
\end{itemize}
\item[(b)] \emph{Functionality change attacks}: Data poisoning attacks that cause a model to learn an unintended functionality;
\item[(c)] \emph{Resource exhaustion attacks} (known as \emph{sponge poisoning attacks}~\cite{DBLP:journals/corr/abs-2203-08147}): Data poisoning attacks that cause exhaustion of resources during system operation;
\item[(d)] \emph{Information embedding attacks}: Data poisoning attacks that embed sensitive information into ML datasets to disclose it during system operation.
\end{enumerate}

Targeted attacks and backdoor attacks~\cite{DBLP:conf/aaai/ChenCBLELMS19,DBLP:journals/corr/abs-2007-08745} are more difficult to be detected by testing the trained model.
This is because these attacks do not cause a malfunction of the trained model except with specific input data or with the input data containing specific triggers.

A functionality change attack can be implemented by replacing the training dataset with another.
For example, a well-known attack~\cite{schwartz2016microsoft} simply adds many data instances containing biased content to a training dataset, resulting in a model that produces discriminatory information.
Since a functionality change attack attempts to train a model with different functionality, it tends to manipulate a large number of training data and may not require prior knowledge of the learning mechanism.
In contrast, the other types of data poisoning attacks may use prior knowledge of the learning mechanism to reduce the amount of manipulation in the dataset.

Information embedding attacks through data poisoning can be categorized into 
(d1) those adding sensitive information to the dataset and 
(d2) those manipulating the dataset to affect the model's behavior during system operation, triggering information leakage attacks of training data (\Sec{sub:V:leakage-data})~\cite{DBLP:conf/sp/MahloujifarGC22,ChaudhariAOJTU:23:SP}.

\subsubsection{Vulnerabilities and Security Controls}
\label{sub:data-poison:VSC}
Data poisoning attacks exploit vulnerabilities of ML data sources (A1, A11), ML datasets (A2, A12), learning mechanisms (A4), and trained models (A5).
In \Tbl{tab:VC:data-poison}, we show the vulnerabilities exploited by data poisoning attacks, and security controls against those vulnerabilities.

\subparagraph{A1, A2, A11, A12: ML data source and ML dataset}
\begin{itemize}
\item
The developers should
evaluate the trustworthiness of the ML data sources (C1.1) and the ML dataset (C2.1).
For example, they should check
the dataset's \emph{authenticity}, e.g., using digital signatures or data provision frameworks;
the data provider's \emph{social credibility};
the \emph{process} of data collection and dataset construction, e.g., concerning the security controls to prevent data manipulation.
However, the developers may not be able to exclude poisoned data instances from the training datasets if they directly collect training data from external environments.

\item
The developers should apply conventional security controls to prevent or mitigate the poisoning of ML data sources and datasets (C1.2, C2.2).
For example, they should apply controls against vulnerabilities in the development software and the development environment, such as software libraries for machine learning (e.g., PyTorch and TensorFlow).

\item
The developers can use techniques to detect the poisoning of ML data sources and datasets (C1.3, C2.3).
For example, a method for detecting outlier data in the training datasets~\cite{DBLP:conf/nips/SteinhardtKL17} effectively removes a small number of poisoned data instances, since the characteristics of poisoned data are close to that of outlier data. 
For another example, the developers may detect functionality change attacks by checking the contents of the dataset automatically or manually, as this type of attack requires manipulating a larger number of data instances in the dataset.

\item
The developers may synthesize or pre-process the ML dataset to make them resilient to data poisoning (C2.4a).
For example, data augmentation~\cite{DBLP:conf/icassp/BorgniaCFGGGGG21} increases the dataset's resilience to poisoning.
Using more training data instances can also mitigate the impact of poisoning without decreasing the accuracy of the trained model.
\end{itemize}

\subparagraph{A4: Learning mechanism}
The resilience to data poisoning attacks also depends on the learning mechanism (C4.1).
For example, the developers may use robust training methods, such as ensemble learning (e.g., Bootstrap Aggregating)~\cite{DBLP:conf/ccs/JiaSBZG19} and randomized smoothing~\cite{DBLP:conf/icml/RosenfeldWRK20}, to mitigate data poisoning.

\subparagraph{A5: Trained model}
The developers can use techniques to remove or reduce poisoning effects from a  trained model (C5.3). 
See \Sec{sub:V:model-poison} for details.

\subparagraph{Survey literature}
For technical details of attacks and defenses, see previous papers, 
e.g.,~\cite{DBLP:journals/corr/abs-2007-08745,DBLP:journals/tse/HeMCHH22,DBLP:journals/corr/abs-2205-01992,wang2022threats,DBLP:journals/corr/abs-2204-05986}.

\subsection{Model Poisoning Attacks}
\label{sub:V:model-poison}

\subsubsection{Overview}
A \emph{model poisoning attack} is an attack that manipulates a pre-trained model, a learning mechanism, or a trained model
to cause the trained model's unintended behavior, the exhaustion of resources by the trained model, or the leakage of sensitive information from the trained model.

\begin{figure}[t]%
\centering
\includegraphics[width=0.98\textwidth]{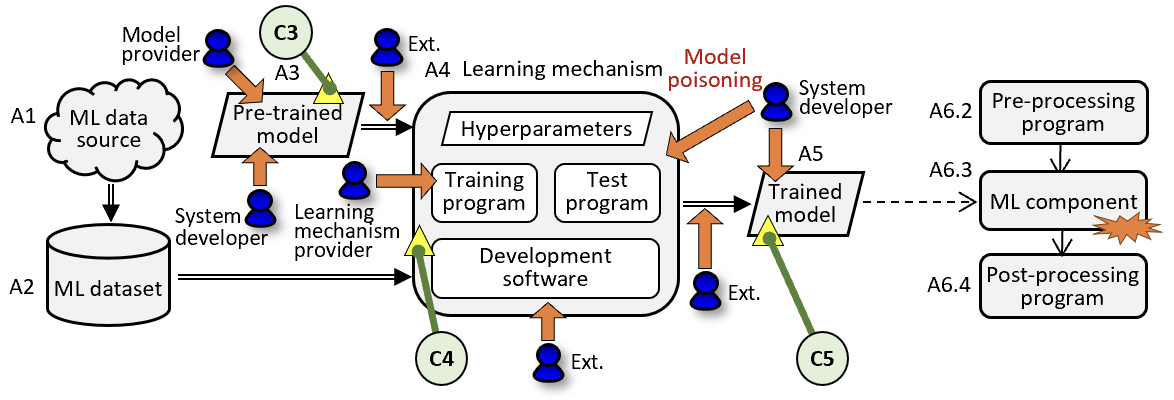}
\caption{Vulnerable assets and security controls against model poisoning attacks.
 Ext denotes an external attacker.}
\label{fig:model-poison}
\end{figure}

\begin{table}[t]
  \caption{Vulnerabilities and controls to \emph{model poisoning attacks} (T1.2).
  See \Tbl{tab:VC:all-threats} for vulnerabilities and controls common to all attack types.}
  \label{tab:VC:model-poison}
\begin{footnotesize}
\begin{tabular}{@{}l@{}l@{}l@{}l@{}l@{}l@{}}\hlineb
\multicolumn{2}{l}{\spB\textgt{Vulnerable asset}} & \multicolumn{2}{l}{\spB\textgt{Vulnerability}} & \multicolumn{2}{l}{\spB\textgt{Control}} \\\hline
A3 & \begin{tabular}{l} Pre-trained \\[-0.3ex] model \end{tabular} & V3.1 & \begin{tabular}{l} Lack of a process to evaluate \\[-0.3ex] the trustworthiness of \\[-0.3ex] pre-trained models \end{tabular} & C3.1 & \begin{tabular}{l} Evaluate the trustworthiness \\[-0.3ex] of pre-trained models \end{tabular} \\
\cline{3-6}
 &  & V3.2 & \begin{tabular}{l} Lack of security controls to \\[-0.3ex] prevent/mitigate the manipulation \\[-0.3ex] of pre-trained models \end{tabular} & C3.2 & \begin{tabular}{l} Apply security controls to \\[-0.3ex] prevent/mitigate the manipulation \\[-0.3ex] of pre-trained models \end{tabular} \\
\cline{3-6}
 &  & V3.3 & \begin{tabular}{l} Lack of a process to detect \\[-0.3ex] poisoning effects from  \\[-0.3ex] pre-trained models \end{tabular} & C3.3 & \begin{tabular}{l} Use techniques to detect \\[-0.3ex] poisoning effects from pre-trained \\[-0.3ex] models \end{tabular} \\
\cline{3-6}
 &  & V3.4 & \begin{tabular}{l} Lack of a process to remove/ \\[-0.3ex]reduce poisoning effects \\[-0.3ex] from pre-trained models \end{tabular} & C3.4 & \begin{tabular}{l} Use techniques to remove/ \\[-0.3ex] reduce poisoning effects from \\[-0.3ex] pre-trained models \end{tabular} \\
\hline
A4 & \begin{tabular}{l} Learning \\[-0.3ex] mechanism \end{tabular} & V4.2 & \begin{tabular}{l} Lack of a process to evaluate \\[-0.3ex] the trustworthiness of learning \\[-0.3ex] mechanisms \end{tabular} & C4.2 & \begin{tabular}{l} Evaluate the trustworthiness \\[-0.3ex] of learning mechanisms \end{tabular} \\
\cline{3-6}
 &  & V4.3 & \begin{tabular}{l} Lack of security controls to \\[-0.3ex] prevent/mitigate the manipulation \\[-0.3ex] of learning mechanisms \end{tabular} & C4.3 & \begin{tabular}{l} Apply security controls to \\[-0.3ex] prevent/mitigate the manipulation \\[-0.3ex] of learning mechanisms \end{tabular} \\
\cline{3-6}
 &  & V4.4 & \begin{tabular}{l} Learning mechanisms not \\[-0.3ex] removing/reducing poisoning \\[-0.3ex] effects from pre-trained models \end{tabular} & C4.4 & \begin{tabular}{l} Use learning mechanisms that \\[-0.3ex] can remove/reduce poisoning \\[-0.3ex] effects from pre-trained models \end{tabular} \\
\hline
A5 & \begin{tabular}{l} Trained \\[-0.3ex] model \end{tabular} & V5.1 & \begin{tabular}{l} Lack of security controls to \\[-0.3ex] suppress/prevent the poisoning \\[-0.3ex] of trained models \end{tabular} & C5.1 & \begin{tabular}{l} Apply security controls to \\[-0.3ex] suppress/prevent the poisoning \\[-0.3ex] of trained models \end{tabular} \\
\cline{3-6}
 &  & V5.2 & \begin{tabular}{l} Lack of a process to detect \\[-0.3ex] poisoning effects in trained models \end{tabular} & C5.2 & \begin{tabular}{l} Use techniques to detect \\[-0.3ex] poisoning effects in trained models \end{tabular} \\
\cline{3-6}
 &  & V5.3 & \begin{tabular}{l} Lack of a process to remove/ \\[-0.3ex] reduce poisoning effects from \\[-0.3ex] trained models \end{tabular} & C5.3 & \begin{tabular}{l} Use techniques to remove/ \\[-0.3ex] reduce poisoning effects from \\[-0.3ex] trained models \end{tabular} \\
\hlineb
\end{tabular}
\end{footnotesize}
\end{table}

\subparagraph{Damage}
This kind of attack is conducted during a pre-trained model's provision, model learning, or model deployment.
The damage caused by this attack occurs during system operation.
As with data poisoning attacks (\Sec{sub:V:data-poison}), 
model poisoning attacks aim to cause a trained model's malfunction, affecting the system’s qualities, such as performance, safety,  fairness, and privacy.

\subparagraph{Attack situation}
We show an overview of the attack situations in model poisoning attacks in \Fig{fig:model-poison}.
Typical attackers and their situations are as follows:
\begin{itemize}
\item An external attacker, a model provider, or a system developer manipulates the pre-trained model (A3), and the poisoning effects in the pre-trained model remain in the trained model after transfer learning.
\item An external attacker, a learning mechanism provider, or a system developer manipulates the learning mechanism (A4). 
The external attacker exploits vulnerabilities of the development software.
The learning mechanism provider can be (i) an external developer to whom the model learning is outsourced or (ii) a development platform for model learning, e.g., Machine Learning as a Service (MLaaS).
\item An external attacker or a system developer manipulates the trained model (A5) in the process of model deployment.
\end{itemize}

\subparagraph{Classification w.r.t. damage}
We list the following major types of model poisoning attacks:
\begin{enumerate}
\item[(a)] \emph{Malfunction attacks}
\begin{itemize}
\item \emph{Targeted attacks}: Model poisoning attacks that cause a malfunction of a trained model for specific inputs to the model;
\item \emph{Backdoor attacks}: Model poisoning attacks that cause a malfunction of a trained model for inputs that contain specific information (triggers);
\item \emph{Non-targeted attacks}: Model poisoning attacks that cause a malfunction of a trained model for unspecified inputs;
\end{itemize}
\item[(b)] \emph{Functionality change attacks}: Model poisoning attacks that cause a model to learn an unintended functionality;
\item[(c)] \emph{Resource exhaustion attacks}: Model poisoning attacks that cause exhaustion of resources during system operation;
\item[(d)] \emph{Information embedding attacks}: Model poisoning attacks that embed sensitive information into model parameters or hyperparameters to disclose them during system operation.
\end{enumerate}

The goals of malfunction attacks through model poisoning are similar to those through data poisoning (\Sec{sub:V:data-poison}).

A functionality change attack via model poisoning can be implemented by replacing the trained model with a different model with unintended functionality.

Information embedding attacks~\cite{DBLP:conf/ccs/SongRS17,DBLP:conf/asiaccs/JiaWG21} exploit a large capacity of a trained model that can encode more information than the model's task.
Sensitive information is embedded into model parameters or hyperparameters before system operation, and is disclosed during system operation.

\subsubsection{Vulnerabilities and Security Controls}
\label{sub:model-poison:VSC}
Model poisoning attacks exploit vulnerabilities of pre-trained models (A3), learning mechanisms (A4), and trained models (A5).
In \Tbl{tab:VC:model-poison}, we show the vulnerabilities exploited by model poisoning attacks, and security controls against those vulnerabilities.

\subparagraph{A3, A5: Pre-trained and trained models}

\begin{itemize}
\item
The developers should evaluate the trustworthiness of the pre-trained model used to train the model (C3.1).
For example, they should check 
the pre-trained model's \emph{authenticity} (e.g., using digital signatures or some model  provision frameworks),
the model provider's \emph{social credibility}, and
the \emph{process} of the training of the pre-trained model (e.g., concerning the security controls in the development environment).
In  practice, however, the developers may not be able to evaluate the pre-trained model sufficiently.

\item
The developers may use techniques to detect poisoning effects from pre-trained models and trained models (C3.3, C5.2).
However, these techniques can fail to detect poisoning or may be evaded by new attack methods.
For example, an attack method~\cite{DBLP:journals/corr/abs-2204-06974} uses cryptographic techniques to install undetectable backdoors into models.

\item
The developers may use techniques to remove or reduce poisoning effects from pre-trained models and trained models, e.g., by model pre-processing, additional learning, and ensemble learning (C3.4, C4.4, C5.3).
For example, a technique~\cite{DBLP:conf/raid/0017DG18} to remove backdoors from deep learning models is to prune some nodes in the neural network and 
update model parameters through additional learning.
Notice that model pre-processing may decrease the model's performance if we do not perform additional learning using sufficient training data.
Another possible technique is ensemble learning, which may mitigate poisoning effects, thanks to other trained models that are not poisoned.

\item
The developers should apply conventional security controls against vulnerabilities in the development environment
to prevent or mitigate the manipulation of the pre-trained model and the trained model (C3.2, C5.1).
\end{itemize}

\subparagraph{A4: Learning mechanism}
The developers should evaluate the trustworthiness of the learning mechanism used to train the model (C4.2).
They should also 
apply conventional security controls against vulnerabilities 
to prevent or mitigate the manipulation of the learning mechanism (C4.3).
For example, they should apply security controls for the development software (e.g., a software library for machine learning) and the development environment.

\subparagraph{Survey literature}
For technical details of attacks and defenses, see previous papers, 
e.g.,~\cite{DBLP:journals/corr/abs-2007-08745,DBLP:journals/tse/HeMCHH22,wang2022threats}.

\subsection{Exploitation of Poisoned Models}
\label{sub:V:EPM}

\subsubsection{Overview}
An \emph{exploitation of a poisoned model} is an attack that inputs malicious data during operation to exploit poisoning 
to cause the trained model's unintended behavior, the exhaustion of resources by the trained model, or the leakage of sensitive information from the trained model.

\begin{figure}[t]%
\centering
\includegraphics[width=0.98\textwidth]{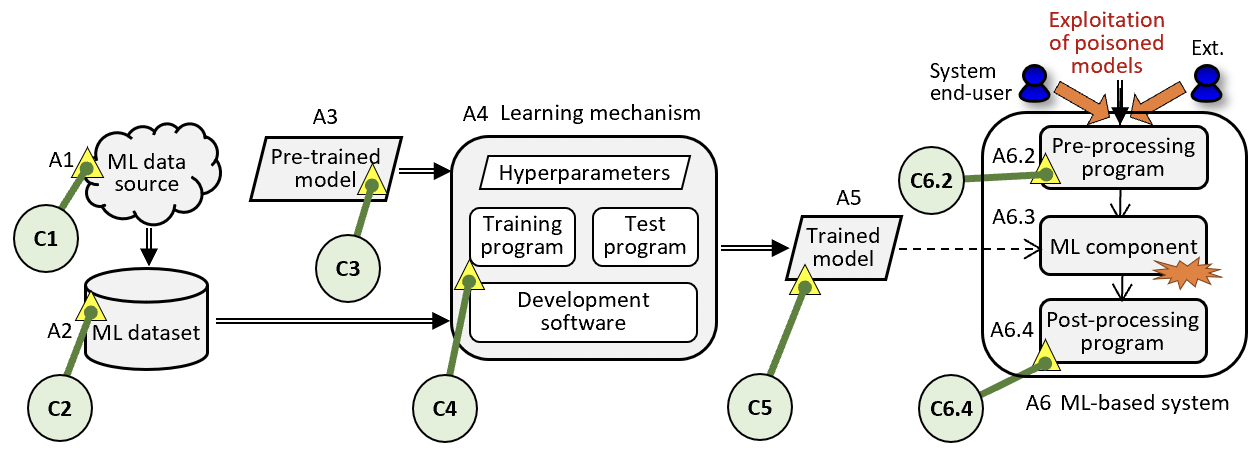}
\caption{Vulnerable assets and security controls against the exploitation of poisoned models.
 Ext denotes an external attacker.}
\label{fig:exploit-poison}
\end{figure}

\begin{table}[t]
  \caption{Vulnerabilities and controls to the \emph{exploitation of a poisoned model} (T2.1) during system operation.
  The symbol \Rare{} indicates that security controls have not been studied sufficiently.
  See \Tbls{tab:VC:all-MIS} and \ref{tab:VC:all-threats} for the vulnerabilities and controls common to other threats.
  }
  \label{tab:VC:EPM}
\begin{footnotesize}
\begin{tabular}{@{}l@{}l@{}l@{}l@{}l@{}l@{}}\hlineb
\multicolumn{2}{l}{\spB\textgt{Vulnerable asset}} & \multicolumn{2}{l}{\spB\textgt{Vulnerability}} & \multicolumn{2}{l}{\spB\textgt{Control}} \\\hline
A6.2 & \begin{tabular}{l} Pre-process-\\[-0.3ex]ing program \end{tabular} & V6.2 & \begin{tabular}{l} Lack of a process \\[-0.3ex] to detect/pre-process/\\[-0.3ex] restrict malicious input \\[-0.3ex] to ML components \\[-0.3ex] during operation \end{tabular} & C6.2 & \begin{tabular}{l} Use techniques to detect/ \\[-0.3ex] pre-process/restrict \\[-0.3ex] malicious input to ML \\[-0.3ex] components during operation \end{tabular} \\
\hline
A6.3 & \begin{tabular}{l} ML \\[-0.3ex] component \end{tabular} & V6.3a & \begin{tabular}{l} ML component that functions \\[-0.3ex] unintendedly, exhausts \\[-0.3ex] resources, or leaks embedded \\[-0.3ex] information for malicious \\[-0.3ex] inputs exploiting the poisoned \\[-0.3ex] model \end{tabular} 
& \begin{tabular}{l} C1 \\ \,to \\ C5 \\~ \end{tabular} & \begin{tabular}{l} Apply security controls to \\[-0.3ex] assets A1 to A5 against \\[-0.3ex] poisoning attacks  \\[-0.3ex] (See \Tbls{tab:VC:data-poison} and \ref{tab:VC:model-poison}) \\[-0.3ex] ~ \end{tabular} \\
\hline
A6.4 & \begin{tabular}{l} Post-process-\\[-0.3ex]ing program \end{tabular} & V6.4 & \begin{tabular}{l} Lack of a process to limit \\ the observation of output \\ and internal information \\ of ML components \\ during operation \end{tabular} & C6.4 \Rare & \begin{tabular}{l} Restrict the disclosure of \\ the output and internal \\ information of ML components \\ during operation \end{tabular} \\
\hlineb
\end{tabular}
\end{footnotesize}
\end{table}

\subparagraph{Damage}
This kind of attack is conducted during system operation.
The damage caused by this attack occurs during system operation.
It is the same as the damage caused by data poisoning attacks (\Sec{sub:V:data-poison}) or model poisoning attacks (\Sec{sub:V:model-poison}).
However, we define this threat separately from the poisoning attacks (conducted \emph{during system development}),
because malicious data input \emph{during system operation} is required to trigger the targeted and backdoor poisoning.

\subparagraph{Attack situation}
We show an overview of the attack situations in the exploitation of poisoned models in \Fig{fig:exploit-poison}.
We assume a situation where a trained model has been poisoned due to data poisoning attacks (\Sec{sub:V:data-poison}) or model poisoning attacks (\Sec{sub:V:model-poison}).
Then a system end-user is assumed to be an attacker that inputs malicious data to the system to cause the system's malfunction, resource exhaustion, or information leakage.
For instance, if a backdoor is embedded into the trained model, the attacker provides the system with input data that trigger the backdoor.

\subsubsection{Vulnerabilities and Security Controls}
The exploitation of a poisoned model exploit vulnerabilities of A6 (the system), A7 (sources of data for system operation), A8 (data for system operation), and A9 (the computing environment and the operating organization).

Among them, in \Tbl{tab:VC:EPM}, we show the vulnerabilities of A6.2 (pre-processing programs), A6.3 (ML components), and A6.4 (post-processing programs), and security controls against those vulnerabilities.
As for the other assets, we will present their vulnerabilities and security controls in \Tbls{tab:VC:all-MIS} and \ref{tab:VC:all-threats} in \Sects{sub:V:all-oracles} and \ref{sub:V:all-threats}, respectively.

\subparagraph{A6.2: Pre-processing program}
The developers can use techniques to detect, pre-process, or restrict malicious input to ML components during operation (C6.2).
\begin{itemize}
\item 
They can use techniques to pre-process the input that may exploit poisoned models.
For example, Februus~\cite{DBLP:conf/acsac/DoanAR20} sanitizes the input during system operation to remove the Trojan trigger that exploits a backdoor embedded into the input to cause the model's malfunction.
\item
Less attention has been paid to detecting malicious inputs to an ML component that exploit a poisoned model during system operation.
For example, MISA~\cite{DBLP:conf/acsac/KiourtiLRSJ21} is an approach to detecting Trojan triggers for neural networks during system operation without requiring any training data injected with Trojan triggers.
\end{itemize}

\subparagraph{A6.3: ML Component}
The essential controls are to prevent or mitigate the poisoning of the trained model by applying security controls C1 to C5 for vulnerabilities V1 to V5 against poisoning itself.
See \Sects{sub:data-poison:VSC} and \ref{sub:model-poison:VSC}.

\subparagraph{A6.4: Post-processing program}
To the best of our knowledge, 
to mitigate this attack,
there have been no practical controls 
to restrict the disclosure of output and internal information of an ML component during operation (C6.4).

\subparagraph{Other assets}
See \Sects{sub:V:all-oracles} and \ref{sub:V:all-threats}, respectively.

\subsection{Model Extraction Attacks}
\label{sub:V:model-extract}

\subsubsection{Overview}
A \emph{model extraction attack} is an attack that inputs malicious data to a trained model during operation 
to cause the leakage of information on the trained model.

\begin{figure}[t]%
\centering
\includegraphics[width=0.98\textwidth]{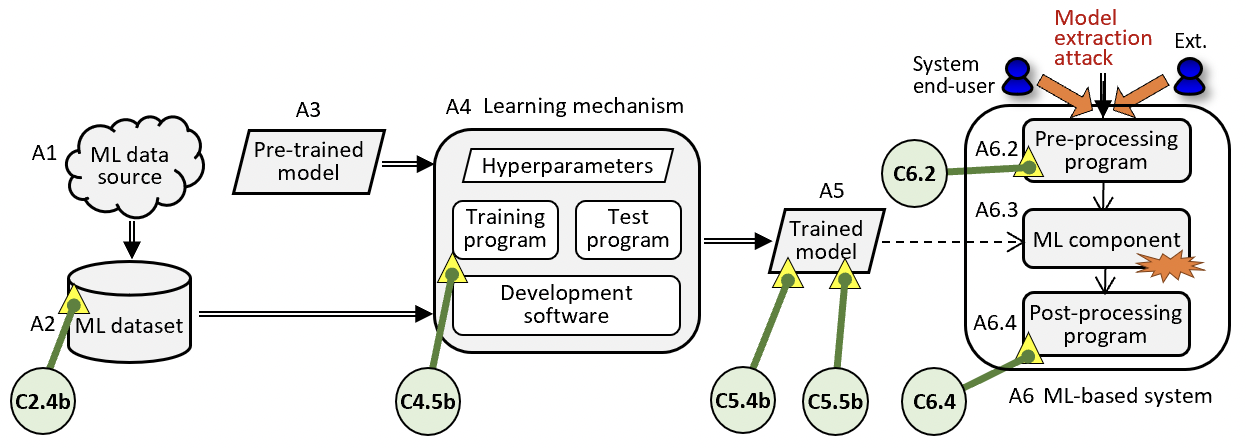}
\caption{Vulnerable assets and security controls against model extraction attacks. 
Ext denotes an external attacker.}
\label{fig:model-extract}
\end{figure}

\begin{table}[t]
  \caption{Vulnerabilities and controls to \emph{model extraction attacks} (T2.2).
  The symbol \Rare{} indicates that security controls have not been studied sufficiently.
  See \Tbls{tab:VC:all-MIS} and \ref{tab:VC:all-threats} for the vulnerabilities and controls common to other threats.}
  \label{tab:VC:model-extract}
\begin{footnotesize}
\begin{tabular}{@{}l@{}l@{}l@{}l@{}l@{}l@{}}\hlineb
\multicolumn{2}{l}{\spB\textgt{Vulnerable asset}} & \multicolumn{2}{l}{\spB\textgt{Vulnerability}} & \multicolumn{2}{l}{\spB\textgt{Control}} \\\hline
A6.2 & \begin{tabular}{l} Pre-process-\\[-0.3ex]ing program \end{tabular} & V6.2 & \begin{tabular}{l} Lack of a process \\[-0.3ex] to detect/pre-process/\\[-0.3ex] restrict malicious input \\[-0.3ex] to ML components \\[-0.3ex] during operation \end{tabular} & C6.2 & \begin{tabular}{l} Use techniques to detect/ \\[-0.3ex] pre-process/restrict \\[-0.3ex] malicious input to ML \\[-0.3ex] components during operation \end{tabular} \\
\hline
A6.3 & \begin{tabular}{l} ML \\[-0.3ex] component \end{tabular} & V6.3b & \multirow{2}{*}{\begin{tabular}{l} ML component that leaks \\ information on a trained \\ model \end{tabular}} & C2.4b \Rare & \begin{tabular}{l} Use techniques to synthesize/ \\ pre-process the ML datasets \\ to mitigate model extraction \end{tabular} \\
\cline{5-6}
 &  &  &  & C4.5b \Rare & \begin{tabular}{l} Use learning mechanisms \\ that produce trained models \\ resilient to their extraction \end{tabular} \\
\cline{5-6}
 &  &  &  & C5.4b & \begin{tabular}{l} Evaluate the risk of extraction \\ of trained models \end{tabular} \\
\cline{5-6}
 &  &  &  & C5.5b \Rare & \begin{tabular}{l} Use techniques to improve \\ trained models to mitigate \\ the leakage of information \\ on trained models \end{tabular} \\
\hline
A6.4 & \begin{tabular}{l} Post-process-\\[-0.3ex]ing program \end{tabular} & V6.4 & \begin{tabular}{l} Lack of a process to limit \\ the observation of output \\ and internal information \\ of ML components \\ during operation \end{tabular} & C6.4 & \begin{tabular}{l} Restrict the disclosure of \\ the output and internal \\ information of ML components \\ during operation \end{tabular} \\
\hlineb
\end{tabular}
\end{footnotesize}
\end{table}

\subparagraph{Damage}
Model extraction attacks may cause the disclosure of trade secrets or other private information on a trained model's attributes and functionality.
Furthermore, the compromised information of the trained model can be used in evasion attacks (\Sec{sub:V:evasion}), sponge attack (\Sec{sub:V:sponge}), and information leakage attacks of training data (\Sec{sub:V:leakage-data}).

\subparagraph{Attack situation}
We show an overview of the attack situations in model extraction attacks in \Fig{fig:model-extract}.
In a model extraction attack, the attacker inputs data to the system during the system operation, and observes the input-output relation of the trained model.
In other words, an attacker (e.g., a system user) is assumed to have black-box access to the trained model under attack.

\subparagraph{Classification w.r.t. damage}
Model extraction attacks are classified into
(i) those to obtain information about a trained model's attributes  (architecture~\cite{DBLP:series/lncs/OhSF19}, hyperparameters~\cite{DBLP:conf/sp/WangG18}, parameters~\cite{DBLP:conf/uss/TramerZJRR16}, decision boundaries~\cite{DBLP:conf/eurosp/JuutiSMA19}, etc.) 
and (ii) those to obtain information about a trained model's functionality~\cite{DBLP:conf/ijcnn/SilvaBBSO18,DBLP:conf/cvpr/OrekondySF19} during the system operation.

\subsubsection{Vulnerabilities and Security Controls}
\label{sub:V:model-extract:VSC}
Model extraction attacks exploit vulnerabilities of A6 (the system), A7 (sources of data for system operation), A8 (data for system operation), and A9 (the computing environment and the operating organization).

Among them, in \Tbl{tab:VC:model-extract}, we show the vulnerabilities of A6.2 (pre-processing programs), A6.3 (ML components), and A6.4 (post-processing programs), and security controls against those vulnerabilities.
As for the other assets, we will present their vulnerabilities and security controls in \Tbls{tab:VC:all-MIS} and \ref{tab:VC:all-threats} in \Sects{sub:V:all-oracles} and \ref{sub:V:all-threats}, respectively.

\subparagraph{A6.2: Pre-processing program}
\begin{itemize}
\item
A helpful security control against this attack is to use techniques to detect and restrict malicious input to an ML component during operation (C6.2).
For instance, the developers can use a tool that observes the distribution of a set of input data to a trained model and detects malicious inputs for model extraction attacks (e.g., PRADA \cite{DBLP:conf/eurosp/JuutiSMA19}).
However, such detection techniques may be less effective when an attacker knows the trained model's task and accesses the training data distribution~\cite{10.1007/978-3-030-62144-5_4}.
\item
There are a few studies on the pre-processing of the input to ML components to mitigate model extraction.
For example, a certain perturbation of input data is empirically shown to reduce the probability of successful model extraction~\cite{DBLP:conf/cns/Grana20}.
\end{itemize}

\subparagraph{A6.3: ML Component}

\begin{itemize}
\item
According to empirical studies (e.g., in \cite{DBLP:conf/uss/LiuWH000CF022}),
using a more complex dataset tends to mitigate model extraction attacks (C2.4b).
To the best of our knowledge, however, this approach has not been studied in terms of defense techniques, and may be potential for future research.

\item
Choices of learning mechanisms may result in different probabilities of successful model extraction (C4.5).
For example, a technique~\cite{DBLP:conf/icissp/ChabanneDG21} to add parasitic layers in neural networks mitigates model extraction.
For another example, empirical experiments~\cite{10.1007/978-3-030-62144-5_4} show that training without using a public pre-trained model may reduce the probability of successful model extraction.
To the best of our knowledge, however, these approaches have not been studied well as defense techniques, and may be potential for future research.

\item
The developers should use evaluation tools for model extraction attacks (C5.4b), such as ML-doctor~\cite{DBLP:conf/uss/LiuWH000CF022}.

\item
To improve a trained model,
the developers may use \emph{ensemble learning} to mitigate the extraction of the trained model by using other multiple models together (C5.5b).
\end{itemize}

\subparagraph{A6.4: Post-processing program}
In designing a post-processing program, the developers may use techniques to restrict the disclosure of the output information of the trained model (C6.4).
In particular, there are techniques to modify the output of the trained model, e.g., 
(i) not outputting confidence values, 
(ii) rounding confidence values, and 
(iii) adding perturbations to confidence values.
While such output modification may mitigate some model extraction attacks, it may not be helpful~\cite{DBLP:conf/eurosp/JuutiSMA19}.
For the mitigation of model extraction, there are differential privacy techniques~\cite{DBLP:conf/esorics/ZhengYHFS19} that add perturbation noise to the prediction responses around the decision boundary.

\subparagraph{Other assets}
See \Sects{sub:V:all-oracles} and \ref{sub:V:all-threats}, respectively.

\subparagraph{Remark on watermarking}
Finally, we remark that \emph{watermarking} \cite{DBLP:conf/mir/UchidaNSS17,DBLP:conf/uss/AdiBCPK18} may be used to protect the intellectual property of a trained model.
Since watermarking does not prevent or mitigate model thefts themselves, it is not included in \Tbl{tab:VC:model-extract}.
However, this technique can be used to trace back the legitimate owner of an already stolen model and may help prevent illegal uses of the stolen model.
For comprehensive surveys, see \cite{DBLP:journals/ijon/LiWB21,DBLP:journals/fdata/Boenisch21}.

\subparagraph{Survey literature}
For technical details of attacks and defenses, see previous papers, 
e.g.,~\cite{DBLP:journals/tse/HeMCHH22,DBLP:journals/corr/abs-2206-08451}.

\subsection{Evasion Attacks}
\label{sub:V:evasion}

\subsubsection{Overview}
An \emph{evasion attack} is an attack that inputs specific malicious data (called \emph{adversarial examples}) to a trained model 
to cause a malfunction of the trained model during operation.
 
\begin{figure}[t]%
\centering
\includegraphics[width=0.98\textwidth]{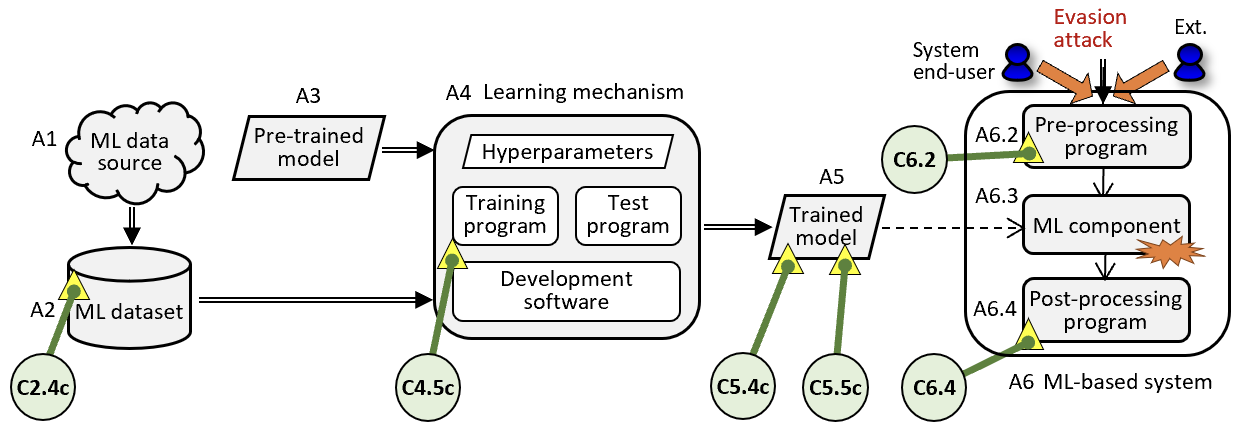}
\caption{Vulnerable assets and security controls against evasion attacks.  Ext denotes an external attacker.}
\label{fig:evasion}
\end{figure}

\subparagraph{Damage}
This kind of attack is conducted during system operation. 
The damage caused by this attack also occurs during system operation.
An evasion attack can malfunction the system to decrease the system's qualities, such as performance and safety. 
For example, in an evasion attack against an image classifier, an adversarial example is a perturbed image input that looks natural to human eyes but is misclassified by the image classifier.

\begin{table}[t]
  \caption{Vulnerabilities and controls to \emph{evasion attacks} (T2.3).
  See \Tbls{tab:VC:all-MIS} and \ref{tab:VC:all-threats} for the vulnerabilities and controls common to other threats.}
  \label{tab:VC:evasion}
\begin{footnotesize}
\begin{tabular}{@{}l@{}l@{}l@{}l@{}l@{}l@{}}\hlineb
\multicolumn{2}{l}{\spB\textgt{Vulnerable asset}} & \multicolumn{2}{l}{\spB\textgt{Vulnerability}} & \multicolumn{2}{l}{\spB\textgt{Control}} \\\hline
A6.2 & \begin{tabular}{l} Pre-process-\\[-0.3ex]ing program \end{tabular} & V6.2 & \begin{tabular}{l} Lack of a process to \\ detect/pre-process/restrict \\ malicious input to ML \\ components during operation \end{tabular} & C6.2 & \begin{tabular}{l} Use techniques to detect/ \\ pre-process/restrict malicious \\ input to ML components \\ during operation \end{tabular} \\
\hline
A6.3 & \begin{tabular}{l} ML \\[-0.3ex] component \end{tabular} & V6.3c & \begin{tabular}{l} ML component that \\ malfunctions on input \\ of adversarial examples  \end{tabular} & C2.4c & \begin{tabular}{l} Use techniques to synthesize/ \\ pre-process the ML datasets \\ to produce robust models \\ against adversarial examples \end{tabular} \\
\cline{5-6}
 &  &  &  & C4.5c & \begin{tabular}{l} Use learning mechanisms \\ that produce robust models \\ against adversarial examples \end{tabular} \\
\cline{5-6}
 &  &  &  & C5.4c & \begin{tabular}{l} Evaluate the robustness of \\ trained models against \\ adversarial examples \end{tabular} \\
\cline{5-6}
  &  &  &  & C5.5c & \begin{tabular}{l} Use techniques to improve the \\ robustness of trained models \\ against adversarial examples \end{tabular} \\
\hline
A6.4 & \begin{tabular}{l} Post-process-\\[-0.3ex]ing program \end{tabular} & V6.4 & \begin{tabular}{l} Lack of a process to limit \\ the observation of output \\ and internal information \\ of ML components \\ during operation \end{tabular} & C6.4 & \begin{tabular}{l} Restrict the disclosure of \\ the output and internal \\ information of ML components \\ during operation \end{tabular} \\

\hlineb
\end{tabular}
\end{footnotesize}
\end{table}

\subparagraph{Requirement on an attacker's knowledge}
Evasion attacks are mainly categorized into \emph{white-box attacks} and \emph{black-box attacks} (\Sec{sub:info:access}).
These two attacks differ in the process and method of generating adversarial examples.

In white-box attacks~\cite{DBLP:conf/iclr/KurakinGB17a,DBLP:journals/access/AkhtarM18}, attackers must obtain a trained model in advance (e.g., by pre-attacks against conventional information systems).
They construct adversarial examples against the trained model by using the model's parameters.
Then they input the adversarial examples into a system that uses the trained model to cause the system's malfunction during its operation.
Since the white-box attackers can directly analyze the model at hand, they do not need to observe the system’s input-output relation during system operation.

In contrast, black-box attacks~\cite{DBLP:conf/ccs/PapernotMGJCS17} generate adversarial examples without using the trained model's parameters.
Typically, to generate adversarial examples, a black-box attacker inputs malicious data to the system during operation and observe the system's input-output relation.
Thus, this kind of attack may be prevented or mitigated by restricting the input data to the system or the attacker's observation of the output.

However, there are blind black-box attacks that do not require multiple input-output pairs of the system.
Such attacks are realized by exploiting the \emph{transferability}~\cite{DBLP:conf/ccs/PapernotMGJCS17} of adversarial examples 
(i.e., the tendency for adversarial examples against a trained model also to be adversarial examples against other trained models).
For example, there are attacks that generate adversarial examples by using
(i) an approximate model that mimics the input-output behavior of the trained model under attack or 
(ii) another model trained using another dataset that resembles the original training dataset.

\subparagraph{Attack situation}
We show an overview of the attack situations in evasion attacks in \Fig{fig:evasion}.
White-box evasion attacks typically assume external attackers that have stolen trained models by exploiting conventional information systems' vulnerabilities.
Black-box evasion attacks typically assume a system end-user as an attacker; 
i.e., a system end-user is assumed to input adversarial examples to the system during operation.
If the system operator uses the system, then an external attacker, the provider of the input data for system operation, or the manager of the data sources can be an attacker.
For example, attackers may manipulate the physical environment (e.g., by printing adversarial example images~\cite{DBLP:conf/iclr/KurakinGB17a,DBLP:conf/cvpr/EykholtEF0RXPKS18} or playing audio adversarial examples~\cite{DBLP:conf/ijcai/YakuraS19}) or the data collection process (e.g., through corrupted devices for image acquisition~\cite{DBLP:conf/cvpr/Moosavi-Dezfooli17}).

\subparagraph{Classification w.r.t. damage}
Evasion attacks are classified in terms of the malfunction of the trained model (\emph{error specificity}~\cite{biggio2018wild}).
\begin{enumerate}
\item[(a)] \emph{Error-generic attacks}: Evasion attacks that cause some unspecified malfunction of the trained model;
\item[(b)] \emph{Error-specific attacks}: Evasion attacks that cause a specific malfunction of the trained model.
\end{enumerate}
In some literature (e.g., in \cite{DBLP:conf/eurosp/PapernotMJFCS16}), (a) is referred to as \emph{non-targeted} and (b) as \emph{targeted}.
\emph{Confidence reduction} is also considered an attack goal to introduce ambiguity in confidence scores.

There is another classification of evasion attacks regarding the range of adversarial inputs during system operation (\emph{attack specificity}).
\begin{enumerate}
\item[(i)] \emph{Indiscriminate attacks}: Evasion attacks that cause malfunctions of trained models for unspecified inputs during system operation;
\item[(ii)] \emph{Targeted attacks}: Evasion attacks that cause malfunctions of trained models for specific inputs during system operation.
\end{enumerate}

\subsubsection{Vulnerabilities and Security Controls}
Evasion attacks exploit vulnerabilities of A6 (the system), A7 (sources of data for system operation), A8 (data for system operation), and A9 (the computing environment and the operating organization).

Among them, in \Tbl{tab:VC:evasion}, we show the vulnerabilities of A6.2 (pre-processing programs), A6.3 (ML components), and A6.4 (post-processing programs), and security controls against those vulnerabilities.
As for the other assets, we will present their vulnerabilities and security controls in \Tbls{tab:VC:all-MIS} and \ref{tab:VC:all-threats} in \Sects{sub:V:all-oracles} and \ref{sub:V:all-threats}, respectively.

Security controls against evasion attacks have been studied primarily in terms of the robustness of trained models against adversarial examples~\cite{DBLP:journals/corr/GoodfellowSS14,DBLP:conf/iclr/KurakinGB17a,DBLP:journals/access/AkhtarM18}.
However, since it is hard to construct perfectly robust models against adversarial examples,
the developers should apply system-level controls, especially restricting the input and output of trained models in the system.

\subparagraph{A6.2: Pre-processing program}
The system should restrict the input to an ML component by detecting, pre-processing, and restricting malicious input to the ML component during operation (C6.2).

\begin{itemize}
\item
The developers can use \emph{adversarial example detection techniques} (e.g.,~\cite{DBLP:journals/air/AldahdoohHFD22,DBLP:conf/ndss/MaLTL019,DBLP:conf/ndss/Xu0Q18}) during system operation.
However, these techniques may fail to detect adversarial examples and can be used only as a secondary measure.

\item
The developers can use \emph{perturbation techniques} to the input during system operation.
For example, they can randomly resize the input or add padding to the input (e.g., \cite{DBLP:conf/iclr/XieWZRY18}).

\item
The developers can use \emph{denoising techniques} to the input during system operation.
Compression-based techniques~\cite{DBLP:journals/corr/DziugaiteGR16,DBLP:conf/cvpr/JiaWCF19} can remove or mitigate adversarial perturbations by compressing the input data.
GAN-based denoising techniques~\cite{DBLP:conf/iclr/SamangoueiKC18} generate images that are close to the original images but do not contain adversarial perturbations by using generative adversarial networks (GANs)~\cite{DBLP:conf/nips/GoodfellowPMXWOCB14}.
Super-resolution-based denoising techniques~\cite{DBLP:journals/tip/MustafaKHSS20} can remove adversarial perturbations while improving task performance thanks to super-resolution.
\end{itemize}

\subparagraph{A6.3: ML Component}
\begin{itemize}
\item The developers can use techniques to synthesize or pre-process the ML datasets  to produce robust models against adversarial examples (C2.4c),
or use learning mechanisms that produce robust models against adversarial examples (C4.5c).
\begin{itemize}
\item
\emph{Adversarial training} is a technique that augments the training dataset with adversarial examples~\cite{DBLP:journals/corr/SzegedyZSBEGF13,DBLP:journals/corr/GoodfellowSS14,DBLP:conf/icdm/LyuHL15,DBLP:conf/iclr/KurakinGB17,DBLP:journals/ijon/ShahamYN18}.
Instead of actually increasing the training dataset itself, the developers can also implement adversarial training by modifying the objective function used in training~\cite{DBLP:journals/corr/GoodfellowSS14}.
Although adversarial training is regarded as the most effective countermeasure against adversarial examples, it may not be sufficient for black-box evasion attacks that generate adversarial examples by using another resembling model~\cite{DBLP:conf/ccs/PapernotMGJCS17,DBLP:conf/cvpr/NarodytskaK17} due to the transferability of adversarial examples~\cite{DBLP:conf/iclr/LiuCLS17}.
To overcome this issue, ensemble adversarial training~\cite{DBLP:conf/iclr/TramerKPGBM18} augments the training dataset with adversarial perturbation transferred from other models.

\item
\emph{High-level representation guided denoiser (HGD)} \cite{DBLP:conf/cvpr/LiaoLDPH018} suppresses the influence of adversarial perturbation by using a loss function that captures the difference between the model's outputs given the original input and those given their corresponding adversarial examples.

\end{itemize}

\item The developers should evaluate the robustness of trained models against adversarial examples (C5.4c).

\begin{itemize}
\item
The evaluation can be performed, e.g.,  
by generating adversarial examples or 
by calculating \emph{maximum safe radius} (i.e., the distance between the original data and their corresponding adversarial examples) exactly~\cite{DBLP:conf/cav/KatzBDJK17,DBLP:conf/iclr/TjengXT19} or approximately~\cite{DBLP:conf/icml/WengZCSHDBD18,DBLP:conf/aaai/BoopathyWC0D19,DBLP:conf/icml/WengCNSBOD19}. 

\item The developers can use publicly available tools for evaluating the robustness of trained models
(e.g., the Adversarial Robustness Toolbox~\cite{nicolae2018adversarial}, RobustBench~\cite{DBLP:conf/nips/CroceASDFCM021}, CleverHans~\cite{papernot2016technical}, and Foolbox~\cite{rauber2017foolbox}).
Since new evaluation tools are developed to cover the latest attack methods,
the developers should find new libraries and benchmarks for the evaluation of the model robustness.
\end{itemize}

\item The developers can
use techniques to improve the robustness of trained models against adversarial examples (C5.5c).
\begin{itemize}
\item
The developers may 
use a \emph{stochastic combination} of multiple models.
For example, hierarchical random switching (HRS)~\cite{DBLP:conf/ijcai/WangWCWKLC19} introduces a chain of random switching blocks that consist of parallel channels and switch them randomly.
The random self-ensemble (RSE) defense method~\cite{DBLP:conf/eccv/LiuCZH18} inserts a layer for adding random noise in the neural network.

\item
Evasion attacks against the interpretation functionality of a trained model~\cite{DBLP:conf/nips/DombrowskiAAAMK19,DBLP:conf/aies/SlackHJSL20} may be prevented or mitigated by using multiple interpretation methods together,
since they tend to be specific to the interpretation method under attack.

\item
\emph{Defensive distillation} \cite{DBLP:conf/sp/PapernotM0JS16} aims to produce a robust model against adversarial examples by using \emph{distillation}~\cite{DBLP:journals/corr/HintonVD15}, that is, a transfer learning method for training a smaller model from a larger original model.
However, owing to the transferability of adversarial examples, certain adversarial examples can evade this defense technique~\cite{DBLP:conf/ccs/PapernotMGJCS17,DBLP:conf/sp/Carlini017}.
\end{itemize}
\end{itemize}
We remark that certain techniques to improve the robustness of trained models may produce models that leak more information about training data
 (\Sec{sub:V:leakage-data})~\cite{DBLP:conf/sp/SongSM19}.

\subparagraph{A6.4: Post-processing program}
The developer should restrict the disclosure of an ML component's output and internal information during operation (C6.4).
This will limit the amount of knowledge that the attacker can extract about the trained model's functionality. 
See \Sec{sub:V:model-extract:VSC} for details.

\subparagraph{Other assets}
See \Sects{sub:V:all-oracles} and \ref{sub:V:all-threats}, respectively.

\subparagraph{Survey literature}
For technical details of attacks and defenses, see previous papers, 
e.g.,~\cite{biggio2018wild,DBLP:journals/caaitrit/ChakrabortyADCM21,ijcai2021p635,electronics11081283,DBLP:journals/air/AldahdoohHFD22,DBLP:journals/access/KhamaisehBAMA22}.

\subsection{Sponge Attacks}
\label{sub:V:sponge}

\subsubsection{Overview}
A \emph{sponge attack} is an attack that inputs malicious data to a trained model 
to cause the exhaustion of resources during system operation.

\begin{figure}[t]%
\centering
\includegraphics[width=0.98\textwidth]{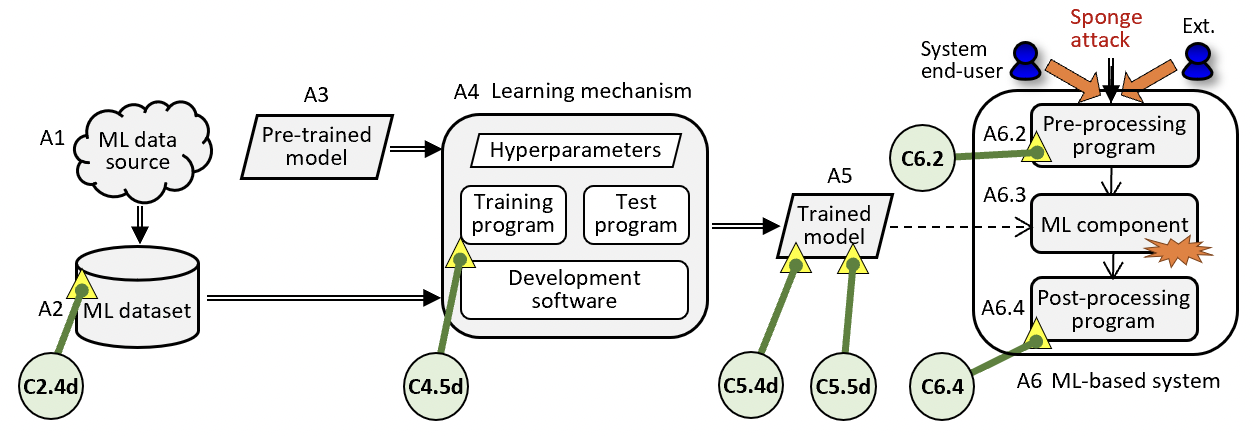}
\caption{Vulnerable assets and security controls against sponge attacks.  Ext denotes an external attacker.}
\label{fig:sponge}
\end{figure}

\begin{table}[t]
  \caption{Vulnerabilities and controls to \emph{sponge attacks} (T2.4).
  The symbol \Rare{} indicates that security controls have not been studied sufficiently.
  See \Tbls{tab:VC:all-MIS} and \ref{tab:VC:all-threats} for the vulnerabilities and controls common to other threats.}
  \label{tab:VC:sponge}
\begin{footnotesize}
\begin{tabular}{@{}l@{}l@{}l@{}l@{}l@{}l@{}}\hlineb
\multicolumn{2}{l}{\spB\textgt{Vulnerable asset}} & \multicolumn{2}{l}{\spB\textgt{Vulnerability}} & \multicolumn{2}{l}{\spB\textgt{Control}} \\\hline
A6.2 & \begin{tabular}{l} Pre-process-\\[-0.3ex]ing program \end{tabular} & V6.2 & \begin{tabular}{l} Lack of a process to \\ detect/pre-process/restrict \\ malicious input to ML \\ components during operation \end{tabular} & C6.2 \Rare & \begin{tabular}{l} Use techniques to detect/ \\ pre-process/restrict malicious \\ input to ML components \\ during operation \end{tabular} \\
\hline
A6.3 & \begin{tabular}{l} ML \\[-0.3ex] component \end{tabular} & V6.3d & \begin{tabular}{l} ML component that exhausts \\ resources by the input of \\ sponge examples \end{tabular} & C2.4d \Rare & \begin{tabular}{l} Use techniques to synthesize/ \\ pre-process the ML datasets \\ to produce robust models \\ against sponge examples \end{tabular} \\
\cline{5-6}
 &  &  &  & C4.5d \Rare & \begin{tabular}{l} Use learning mechanisms \\ that produce robust models \\ against sponge examples \end{tabular} \\
\cline{5-6}
 &  &  &  & C5.4d \Rare & \begin{tabular}{l} Evaluate the robustness of \\ trained models  against \\ sponge examples  \end{tabular} \\
\cline{5-6}
 &  &  &  & C5.5d \Rare & \begin{tabular}{l} Use techniques to improve the \\ robustness of trained models \\ against sponge examples \end{tabular} \\
\hline
A6.4 & \begin{tabular}{l} Post-process-\\[-0.3ex]ing program \end{tabular} & V6.4 & \begin{tabular}{l} Lack of a process to limit \\ the observation of output \\ and internal information \\ of ML components \\ during operation \end{tabular} & C6.4 \Rare & \begin{tabular}{l} Restrict the disclosure of the \\ output and internal information \\ of ML components during \\ operation \end{tabular} \\
\hlineb
\end{tabular}
\end{footnotesize}
\end{table}

\subparagraph{Damage}
This kind of attack is conducted during system operation. 
The damage caused by this attack also occurs during system operation.
A sponge attack can exhaust computational resources to compromise their availability, resulting in decreases in the system's qualities, such as performance and safety.
For example, in a sponge attack against a deep neural network~\cite{DBLP:conf/eurosp/ShumailovZBPMA21}, an attacker designs input data (called \emph{sponge examples}) to  maximize the energy consumption and latency of the ML component to cause a denial of service of the system and possibly result in slow decisions in safety-critical systems.

\subparagraph{Requirement on an attacker's knowledge}
As with evasion attacks, sponge attacks are also categorized into white-box attacks and black-box attacks (\Sec{sub:info:access}).
Unlike evasion attacks, an interactive black-box attack allows an attacker to measure the time and energy consumption of the ML component during operation.
This attack type may not require the attacker's prior knowledge of the model architecture or the dataset used to train the model, as demonstrated in~\cite{DBLP:conf/eurosp/ShumailovZBPMA21}.
Analogously to evasion attacks, attackers may mount a blind black box attack by exploiting the \emph{transferability} of sponge examples (i.e., the tendency for sponge examples against a trained model also to be those against other trained models).

\subparagraph{Attack situation}
We show an overview of the attack situations in sponge attacks in \Fig{fig:sponge}.
White-box sponge attacks typically assume external attackers that steal trained models by exploiting conventional information systems' vulnerabilities.
Black-box sponge attacks typically assume a system end-user as an attacker; 
i.e., we assume a situation where a system end-user inputs sponge examples to the system during operation.
When the system operator uses the system, the provider of the input data for system operation is assumed to be an attacker.

\subparagraph{Remark on sponge poisoning}
We remark that, as mentioned in \Sec{sub:V:data-poison}, a sponge poisoning attack~\cite{DBLP:journals/corr/abs-2203-08147} is a data poisoning attack that causes exhaustion of resources during system operation.
This attack is conducted during system development, and its damage occurs during system operation.

\subsubsection{Vulnerabilities and Security Controls}
Sponge attacks exploit vulnerabilities of A6 (the system), A7 (sources of data for system operation), A8 (data for system operation), and A9 (the computing environment and the operating organization).

Among them, in \Tbl{tab:VC:sponge}, we show the vulnerabilities of A6.2 (pre-processing programs), A6.3 (ML components), and A6.4 (post-processing programs), and security controls against those vulnerabilities.
As for the other assets, we will present their vulnerabilities and security controls in \Tbls{tab:VC:all-MIS} and \ref{tab:VC:all-threats} in \Sects{sub:V:all-oracles} and \ref{sub:V:all-threats}, respectively.

A practical control is to monitor and limit the maximum consumption of energy and other resources, which we mention as C6.5 in \Sec{sub:V:all-threats}.
Since the first paper~\cite{DBLP:conf/eurosp/ShumailovZBPMA21} on sponge attacks was published in 2021, the security controls C6.2, C2.4d, C4.5d, C5.4d, C5.5d, C6.4 in \Tbl{tab:VC:sponge} are still the areas for possible future research.

\subsection{Information Leakage Attacks of Training Data}
\label{sub:V:leakage-data}

\subsubsection{Overview}
An \emph{information leakage attack of training data} is an attack that inputs malicious data to a trained model during operation 
to cause the leakage of sensitive information in a training dataset used to train the model.

\subparagraph{Damage}
This kind of attack is conducted during system operation. 
The damage caused by this attack also occurs during system operation.
An information leakage attack of training data can leak sensitive information in a training dataset to compromise their confidentiality, resulting in the leakage of personal information, trade secrets, or other confidential information.
For example, 
by observing a trained model’s behavior during system operation,
an attacker attempts to infer whether an individual's data record is included in a training dataset (membership inference~\cite{DBLP:conf/sp/ShokriSSS17}).

\subparagraph{Requirement on an attacker's knowledge}
There are two types of information leakage attacks of training data: \emph{white-box attacks} and \emph{black-box attacks} (\Sec{sub:info:access}).
In white-box attacks, attackers must obtain a trained model in advance.
Since the attackers can directly observe the model’s behavior at hand, they do not need to observe the system’s input or output during system operation.
In contrast, black-box attacks assume that attackers can input malicious data into the system during operation and observe the system’s input-output relation.

\begin{figure}[t]%
\centering
\includegraphics[width=0.98\textwidth]{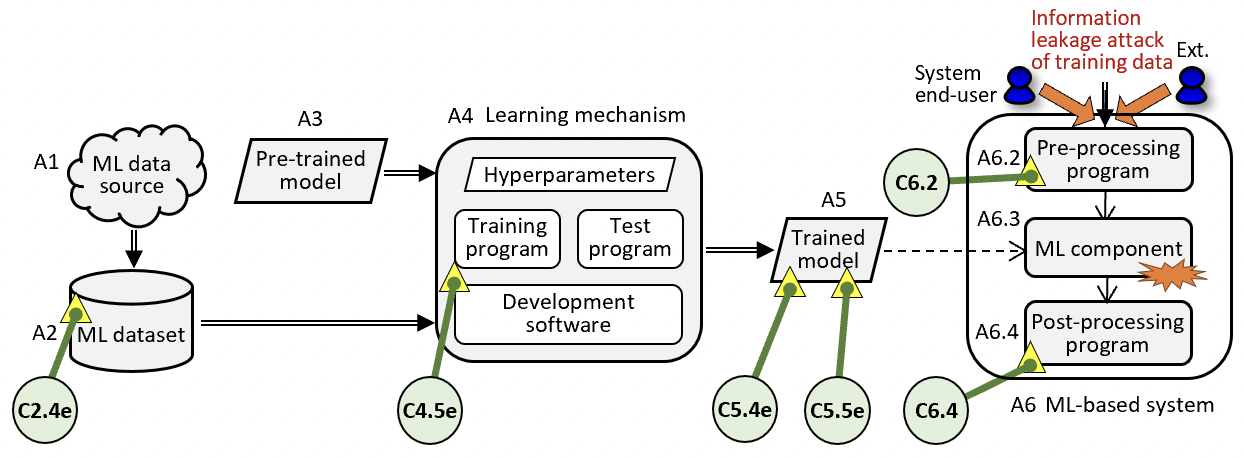}
\caption{Vulnerable assets and security controls against information leakage attacks of training data.  Ext denotes an external attacker.}
\label{fig:data-leak}
\end{figure}

\begin{table}[t]
  \caption{Vulnerabilities and controls to \emph{information leakage attacks of training data} (T2.5).
  The symbol \Rare{} indicates that security controls have not been studied sufficiently.
  See \Tbls{tab:VC:all-MIS} and \ref{tab:VC:all-threats} for the vulnerabilities and controls common to other threats.}
  \label{tab:VC:leakage-data}
\begin{footnotesize}
\begin{tabular}{@{}l@{}l@{}l@{}l@{}l@{}l@{}}\hlineb
\multicolumn{2}{l}{\spB\textgt{Vulnerable asset}} & \multicolumn{2}{l}{\spB\textgt{Vulnerability}} & \multicolumn{2}{l}{\spB\textgt{Control}} \\\hline
A6.2 & \begin{tabular}{l} Pre-process-\\[-0.3ex]ing program \end{tabular} & V6.2 & \begin{tabular}{l} Lack of a process to \\ detect/pre-process/restrict \\ malicious input to ML \\ components during operation \end{tabular} & C6.2 \Rare{} & \begin{tabular}{l} Use techniques to detect/ \\ pre-process/restrict malicious \\ input to ML components \\ during operation \end{tabular} \\
\hline
A6.3 & \begin{tabular}{l} ML \\[-0.3ex] component \end{tabular} & V6.3e & \begin{tabular}{l} ML component that leaks \\ sensitive information \\ on training data \end{tabular} & C2.4e & \begin{tabular}{l} Use techniques to synthesize/ \\ pre-process the ML datasets \\ to mitigate the leakage of \\ sensitive information in the \\ training datasets \end{tabular} \\
\cline{5-6}
 &  &  &  & C4.5e & \begin{tabular}{l} Use learning mechanisms \\ that can prevent/mitigate \\ the leakage of sensitive \\ information in a training dataset \end{tabular} \\
\cline{5-6}
 &  &  &  & C5.4e & \begin{tabular}{l} Evaluate the risk of information \\ leakage from trained models \end{tabular} \\
\cline{5-6}
 &  &  &  & C5.5e & \begin{tabular}{l} Use techniques to improve  \\ trained models to mitigate \\ the leakage of information \\ in the training dataset \end{tabular} \\
\hline
A6.4 & \begin{tabular}{l} Post-process-\\[-0.3ex]ing program \end{tabular} & V6.4 & \begin{tabular}{l} Lack of a process to limit \\ the observation of output \\ and internal information \\ of ML components \\ during operation \end{tabular} & C6.4 & \begin{tabular}{l} Restrict the disclosure of \\ the output and internal \\ information of ML components \\ during operation \end{tabular} \\
\hlineb
\end{tabular}
\end{footnotesize}
\end{table}

\subparagraph{Attack situation}

We show an overview of the attack situations in information leakage attacks of training data in \Fig{fig:data-leak}.

In information leakage attacks of training data against an ML-based system (threat T2.5),
a typical white-box attacker is an external attacker that has stolen trained models by exploiting conventional information systems' vulnerabilities in advance;
a typical black-box attacker is a system end-user,
i.e., a system end-user inputs malicious data to the system during operation to obtain sensitive information in the training dataset used to train the model.
When the system operator uses the system, the provider of the input data for system operation is assumed to be an attacker.

In information leakage attacks of training data against an ML component (threat T3), 
an external third party uses a trained model
and is assumed to be a white-box attacker that attempts to obtain sensitive information in the training dataset used to train the model.

\subparagraph{Classification w.r.t. leaked information}
Information leakage attacks of training data include the following sub-categories of attacks:
\begin{itemize}
\item \emph{Membership inference attacks}~\cite{DBLP:conf/sp/ShokriSSS17} 
attempt to infer whether a specific data record belongs to the training dataset used to train the model.
\item \emph{Attribute inference attacks}~\cite{DBLP:conf/ccs/FredriksonJR15,DBLP:conf/csfw/0001FJN16,DBLP:conf/csfw/YeomGFJ18}
attempt to infer a training data record's sensitive attribute from given other partial information on the record.
\emph{Model inversion attacks}~\cite{DBLP:conf/ccs/FredriksonJR15,DBLP:conf/csfw/0001FJN16}
are regarded as variants of attribute inference attacks that infer a sensitive attribute of a record in a training data \emph{distribution} rather than in a training dataset itself~\cite{DBLP:conf/csfw/YeomGFJ18,DBLP:journals/corr/abs-2212.10986}.
\item \emph{Data reconstruction attacks}~\cite{DBLP:conf/uss/Carlini0EKS19,DBLP:conf/uss/CarliniTWJHLRBS21,DBLP:conf/uss/000100S022,DBLP:conf/sp/BalleCH22} attempt to reconstruct an entire training data record in the training dataset used to train the model.
\item \emph{Property inference attacks}~\cite{DBLP:conf/ccs/GanjuWYGB18,DBLP:conf/ccs/PasquiniAB21,DBLP:conf/uss/0001TO21,DBLP:journals/corr/abs-2207-08367} attempt to infer global properties about the training dataset.
\end{itemize}

We emphasize that information leakage attacks of training data are not limited to the above sub-categories.
Since new attack methods with new attack goals have been actively proposed and studied, this paper does not go into detail and collectively refers to these types of attacks as \emph{information leakage attacks of training data}.
For details, see previous papers on surveys and taxonomies~\cite{DBLP:journals/corr/abs-2007-07646,DBLP:journals/corr/abs-2107-01614,DBLP:journals/corr/abs-2212.10986}.

Finally, we remark that certain poisoning attacks during system development aim to leak information in the training dataset during system operation.
Information embedding attacks through data poisoning (\Sec{sub:V:data-poison}) can trigger information leakage attacks of training data during system operation~\cite{DBLP:conf/sp/MahloujifarGC22,ChaudhariAOJTU:23:SP}.
Information embedding attacks through model poisoning (\Sec{sub:V:model-poison})  embeds sensitive information in advance to disclose it during system operation~\cite{DBLP:conf/ccs/SongRS17,DBLP:conf/asiaccs/JiaWG21}.

\subsubsection{Vulnerabilities and Security Controls}
Information leakage attacks of training data exploit vulnerabilities of A6 (the system), A7 (sources of data for system operation), A8 (data for system operation), and A9 (the computing environment and the operating organization).

Among them, in \Tbl{tab:VC:leakage-data}, we show the vulnerabilities of A6.2 (pre-processing programs), A6.3 (ML components), and A6.4 (post-processing programs), and security controls against those vulnerabilities.
As for the other assets, we will present their vulnerabilities and security controls in \Tbls{tab:VC:all-MIS} and \ref{tab:VC:all-threats} in \Sects{sub:V:all-oracles} and \ref{sub:V:all-threats}, respectively.

\subparagraph{A6.2: Pre-processing program}
To prevent or mitigate information leakage attacks of training data,
the developers may be able to develop techniques to detect, pre-process, and restrict malicious input to ML components during operation (C6.2).
For example, they can apply techniques to detect model extraction attacks (\Sec{sub:V:model-extract:VSC}) so that the attacker fails to learn the trained model's behavior.
Since there are various types of attack algorithms for information leakage attacks of training data, the development of practical detection techniques may be potential for future research.

\subparagraph{A6.3: ML Component}
\begin{itemize}
\item The developers can use techniques to synthesize or pre-process the ML datasets to mitigate the leakage of sensitive information in the training datasets (C2.4e).
For example, they can use the training data generated by privacy-preserving data synthesis techniques~\cite{tucker2020generating}.
The developers may be able to remove or reduce sensitive information from the ML datasets or augment the ML datasets with new data to reduce the impact of these sensitive data.
For instance, they may apply data obfuscation techniques to perturb sensitive information in the data by adding noise to the data (e.g., data obfuscation~\cite{DBLP:journals/corr/abs-1807-01860} using differential privacy~\cite{DBLP:conf/icalp/Dwork06}).

\item The developers can use learning mechanisms that can prevent or mitigate the leakage of sensitive information in a training dataset (C4.5e).
For example, they may use techniques to add differentially private noise to gradient descent computations~\cite{DBLP:conf/focs/BassilyST14,DBLP:conf/globalsip/SongCS13,DBLP:conf/ccs/ShokriS15,DBLP:conf/ccs/AbadiCGMMT016}.

\item The developers can evaluate the risk of information leakage from trained models (C5.4e).
For example, they may use tools to evaluate information leakage of training data from the trained model.
Well-known tools are ML Privacy Meter~\cite{DBLP:journals/corr/abs-2007-09339} and ML-Doctor~\cite{DBLP:conf/uss/LiuWH000CF022}.

\item The developers can 
use techniques to improve trained models to mitigate the leakage of information in the training dataset (C5.5e).
For example, \emph{machine unlearning}~\cite{DBLP:conf/sp/CaoY15,DBLP:conf/cvpr/GolatkarAS20,DBLP:journals/ml/BaumhauerSZ22,DBLP:journals/corr/abs-2209-02299} is a technique to make a trained model forget about particular data.
For another example, \emph{knowledge distillation} is a technique to compress trained models, but can also be used to prevent or mitigate information leakage attacks of training data;
e.g., \emph{distillation for membership privacy} (DMP)~\cite{DBLP:conf/aaai/ShejwalkarH21} leverages knowledge distillation to train a model with membership privacy by adding noise in the process of knowledge distillation.
\end{itemize}

\subparagraph{A6.4: Post-processing program}
The developers should restrict the disclosure of the output and internal information of ML components during operation (C6.4).
For example, \emph{confidence score masking} is a technique to reduce the information in the confidence scores of the model's outputs,
e.g., by providing top-$k$ confidence scores~\cite{DBLP:conf/sp/ShokriSSS17}
or only prediction labels~\cite{DBLP:conf/icml/Choquette-ChooT21}.
However, these na\"{i}ve approaches may not work for some attacks.
MemGuard~\cite{DBLP:conf/ccs/JiaSBZG19} is a technique that adds adversarial-examples-based noises to the confidence score vectors of the trained model.
This method is useful for the defense against DNN-based attackers, but may not work against other attackers~\cite{DBLP:conf/uss/SongM21}.

\subparagraph{Other assets}
See \Sects{sub:V:all-oracles} and \ref{sub:V:all-threats}, respectively.

\subparagraph{Survey literature}
For technical details of attacks and defenses, see previous papers, 
e.g.,~\cite{DBLP:journals/corr/abs-2007-07646,
DBLP:journals/corr/abs-2103-07853,
DBLP:journals/access/TanuwidjajaCBK20,
DBLP:journals/corr/abs-2004-12254,
DBLP:journals/corr/abs-2107-01614,
DBLP:journals/csur/LiuDSRFL21,
DBLP:journals/corr/abs-2212.10986}
~\\

Finally, this paper focuses on centralized (supervised) learning and does not deal with distributed learning.
As for the information leakage attacks in federated learning, see, e.g., \cite{DBLP:conf/sp/NasrSH19,DBLP:journals/ftml/KairouzMABBBBCC21,DBLP:journals/cacm/BonawitzKMR22}

\subsection{System-Level Vulnerabilities and Controls to Malicious Input to Systems}
\label{sub:V:all-oracles}

\begin{figure}[t]%
\centering
\includegraphics[width=0.99\textwidth]{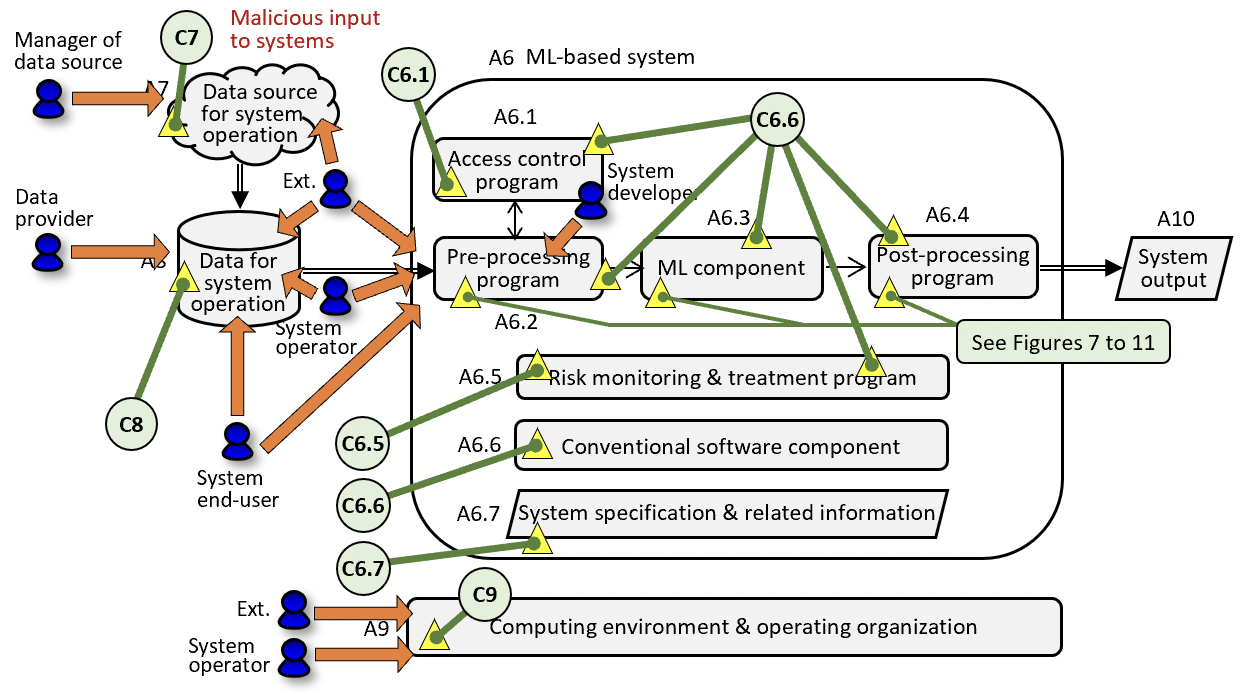}
\caption{Vulnerable assets and security controls during system operation. Ext denotes an external attacker.}
\label{fig:atk-system}
\end{figure}

\begin{table}[t]
  \caption{Vulnerabilities and controls to \emph{all types of threats during system operation} (T2.$*$), i.e., the \emph{input of malicious data to systems}.
  See \Tbl{tab:VC:all-threats} for vulnerabilities and controls common to all attack types during system development and operations.}
  \label{tab:VC:all-MIS}
\begin{footnotesize}
\begin{tabular}{@{}l@{}l@{}l@{}l@{}l@{}l@{}}\hlineb
\multicolumn{2}{l}{\spB\textgt{Vulnerable asset}} & \multicolumn{2}{l}{\spB\textgt{Vulnerability}} & \multicolumn{2}{l}{\spB\textgt{Control}} \\\hline
A6.1 & \begin{tabular}{l} Access control \\ program \end{tabular} & V6.1 & \begin{tabular}{l} Lack of proper access control \\ for ML components during \\ operation \end{tabular} & C6.1 & \begin{tabular}{l} Enforce access control for \\ ML components during \\ operation \end{tabular} \\
\hline
A7 & \begin{tabular}{l} Data source  \\ for system \\ operation \end{tabular} & V7.1 & \begin{tabular}{l} Lack of a process to evaluate \\ the trustworthiness of sources \\ of data for system operation \end{tabular} & C7.1 & \begin{tabular}{l} Evaluate the trustworthiness \\ of sources of data for system \\ operation \end{tabular} \\
\cline{3-6}
 &  & V7.2 & \begin{tabular}{l} Lack of security controls \\ to prevent/mitigate the \\ manipulation of sources of \\ data for system operation \end{tabular} & C7.2 & \begin{tabular}{l} Apply security controls \\ to prevent/mitigate the \\ manipulation of sources of \\ data for system operation \end{tabular} \\
\cline{3-6}
 &  & V7.3 & \begin{tabular}{l} Lack of a process to detect \\ the manipulation of sources \\ of data for system operation \end{tabular} & C7.3 & \begin{tabular}{l} Use techniques to detect \\ the manipulation of sources \\ of data for system operation \end{tabular} \\
\hline
A8 & \begin{tabular}{l} Data for system \\ operation \end{tabular} & V8.1 & \begin{tabular}{l} Lack of a process to evaluate \\ the trustworthiness of data \\ for system operation \end{tabular} & C8.1 & \begin{tabular}{l} Evaluate the trustworthiness \\ of data for system operation \end{tabular} \\
\cline{3-6}
 &  & V8.2 & \begin{tabular}{l} Lack of security controls \\ to prevent/mitigate the \\ manipulation of data for \\ system operation \end{tabular} & C8.2 & \begin{tabular}{l} Apply security controls \\ to prevent/mitigate the \\ manipulation of data for \\ system operation \end{tabular} \\
\cline{3-6}
 &  & V8.3 & \begin{tabular}{l} Lack of a process to detect \\ the manipulation of data for \\ system operation \end{tabular} & C8.3 & \begin{tabular}{l} Use techniques to detect \\ the manipulation of data \\ for system operation \end{tabular} \\
\hlineb
\end{tabular}
\end{footnotesize}
\end{table}

Each ML-specific threat during system operation may exploit vulnerabilities of A6.1 (access control programs), A7 (sources of data for system operation), and A8 (data for system operation).
As discussed in \Sec{sub:attack-surface}, the attack surface during system operation is the assets involved in the input data to the system.
In \Fig{fig:atk-system}, we show an overview of vulnerable assets and security controls during system operation.

In \Tbl{tab:VC:all-MIS}, we briefly show the system-level vulnerabilities common in all types of threats during system operation (T2.$*$), i.e., the input of malicious data to systems.

\subparagraph{A6.1: Access control program}
The system should enforce access control for ML components during operation.
For example, it should restrict the number and frequency of accesses to an ML component and the access rights to the system (C6.1).
This control will decrease: 
\begin{itemize}
\item
the number of malicious inputs to the system for 
the exploitation of poisoned models (\Sec{sub:V:EPM}), evasion attacks (\Sec{sub:V:evasion}), sponge attacks (\Sec{sub:V:sponge}), and information leakage attacks of training data (\Sec{sub:V:leakage-data});
\item
the number of input data that attempt to extract the functionality of the trained model during system operation.
\end{itemize}

\subparagraph{A7, A8: Data and data source for system operation}
The vulnerabilities of data and data sources for system operation (V7.1 to V7.3 and V8.1 to V8.3) are analogous to those for ML datasets and ML data sources (V1.1 to V1.3 and V2.1 to V2.3), shown in \Tbl{tab:VC:data-poison}.
For example, attackers in physical worlds may manipulate (a source of) data for system operation to contaminate it with adversarial examples \cite{DBLP:conf/iclr/KurakinGB17a}. 

Thus, the developers should ensure the adequacy of data and data sources for system operation (C7.$*$, C8.$*$).
This control will prevent the contamination of input with malicious data that cause attacks during system operation.

\subsection{System-Level Vulnerabilities and Controls to Common in ML-Specific Threats}
\label{sub:V:all-threats}

\begin{table}[t]
  \caption{Vulnerabilities and controls of ML-based systems during operation \emph{common in ML-specific threats} (T$*$).}
  \label{tab:VC:all-threats}
\begin{footnotesize}
\begin{tabular}{@{}l@{}l@{}l@{}l@{}l@{}l@{}}\hlineb
\multicolumn{2}{l}{\spB\textgt{Vulnerable asset}} & \multicolumn{2}{l}{\spB\textgt{Vulnerability}} & \multicolumn{2}{l}{\spB\textgt{Control}} \\\hline
A6.5 & \begin{tabular}{l} Monitoring/\\ risk treatment \\ program \end{tabular} & V6.5 & \begin{tabular}{l} Lack of security controls to \\ monitor the system's \\ behavior and treat the risks \\ caused by ML components \end{tabular} & C6.5 & \begin{tabular}{l} Apply security controls to \\ monitor the system's \\ behavior and treat the risks \\ caused by ML components
 \end{tabular} \\
\hline
A6.1-A6.5 & \begin{tabular}{l} Programs directly \\ supporting the \\ model operation \end{tabular} & V6.6 & \begin{tabular}{l} Vulnerability of conventional \\ software \end{tabular} & C6.6 & \begin{tabular}{l} Apply security controls \\ for the vulnerability of \\ conventional software \end{tabular} \\
\hline
A6.6 & \begin{tabular}{l} Other \\ conventional \\ software \\ components \end{tabular} & V6.6 & \begin{tabular}{l} Vulnerability of conventional \\ software \end{tabular} & C6.6 & \begin{tabular}{l} Apply security controls \\ for the vulnerability of \\ conventional software \end{tabular} \\
\hline
A6.7 & \begin{tabular}{l} System \\ specification, \\ etc. \end{tabular} & V6.7 & \begin{tabular}{l} Not restricting the disclosure \\ of the ML datasets, the \\ trained models, the other \\ system specifications, or \\ their related information \end{tabular} & C6.7 & \begin{tabular}{l} Restrict the disclosure of \\ the ML datasets, the \\ trained  models, the other \\ system specifications, and \\ their related information \end{tabular} \\
\hline
A9 & \begin{tabular}{l} Computing \\ environment \\ \& operation \\ organization \end{tabular} & V9.1 & \begin{tabular}{l} Vulnerabilities of computing \\ environment and operation \\ organization during system \\ operation \end{tabular} & C9.1 & \begin{tabular}{l} Apply security controls \\ for the vulnerability of the \\ computing environment and \\ the operating organization \\ during system operation \end{tabular} \\
\cline{3-6}
 &  & V9.2 & \begin{tabular}{l} Lack of continuous updates \\ of security controls by the \\ system operator to cope \\ with the changes in the \\ system and the environment \end{tabular} & C9.2 & \begin{tabular}{l} Update security controls \\ continuously by the system \\ operator to cope with \\ the changes in the system \\ and the environment \end{tabular} \\
\cline{3-6}
 &  & V9.3 & \begin{tabular}{l} Lack of monitoring by the \\ system operator concerning \\ attacks and damage \end{tabular} & C9.3 & \begin{tabular}{l} Enable the system operator \\ to monitor attacks and \\ damage manually  \end{tabular} \\
\hlineb
\end{tabular}
\end{footnotesize}
\end{table}

Each ML-specific threat may exploit vulnerabilities of the system or the operating organization.
See \Fig{fig:atk-system} for an overview of vulnerable assets and security controls during system operation.

In \Tbl{tab:VC:all-threats}, we briefly present the system-level vulnerabilities of ML-based systems common in ML-specific threats.

\subparagraph{A6.5: Monitoring/risk treatment program}
As mentioned in \Sec{sub:damage}, a system malfunction can be caused by the unintended behavior of an ML component or the exhaustion of resources by an ML component.
Therefore, the system should monitor the behavior of the ML components and the entire system, and treat the risks caused by the ML components (C6.5).
\begin{itemize}
\item
\emph{ML component monitoring}:
The system may use explainability techniques in the monitoring of the ML components deployed in the system~\cite{DBLP:conf/fat/BhattXSWTJGPME20}.
For example, the system operators may use them to figure out how and why a specific input caused a model malfunction.
For another example, the system may need to detect data drift, i.e., to check whether the input distribution during system operation has diverged from the training data distribution.

\item
\emph{Resource monitoring}:
A reasonable security control against sponge attacks is to monitor the  computation time, energy consumption, and other resource usages in the system during the system operation,
and to terminate computations beyond cut-off thresholds~\cite{DBLP:conf/eurosp/ShumailovZBPMA21,DBLP:journals/corr/abs-2203-08147}.
\end{itemize}

\subparagraph{A6.6, A9: Other conventional software components, computing environment, and operating organization}
As discussed in \Sec{sub:pre-attack},
ML-specific threats may rely on pre-attacks that exploit conventional vulnerabilities,
namely those of conventional software components in the system (V6.6) and
those of the computing environment and the operating organization (V9.1).
Therefore, the system operators should collect the latest information on vulnerabilities
and implement conventional security controls for those vulnerabilities (C6.6, C9.1).

ML-specific threats may take advantage of the lack of updates of security controls by the system operator (V9.2).
Therefore, the system operators should update security controls continuously to cope with the changes in the system and the environment (C9.2).
\begin{itemize}
\item 
For example, to cope with the decrease in the model's quality during operation (e.g., by data drift~\cite{tsymbal2004problem}) or to recover from the damage by attacks, the system operators may need to rewind the model to a previous version or re-train the model.
In the case of re-training, the stakeholders should enforce the same security controls as those for model training during the development phase.
\item 
For another example, the system operators may change the system's external environment (e.g., location) to reduce the opportunity for malicious input to the system.
\end{itemize}

ML-specific threats may also take advantage of the lack of monitoring by the system operator (V9.3).
Therefore, the system may need to enable the system operators to monitor attacks and damage manually (C9.3).
In many systems, however, the operator's manual confirmation is hard or expensive.

\subparagraph{A6.7: System specification, etc}

As discussed in \Sec{sub:attacker}, ML-specific attacks become feasible or more efficient
if attackers can access more information on the ML-based system under attack (V6.7).
Therefore, the developers should apply controls to restrict the disclosure of the ML datasets, the trained models, the other specification, and their related information (C6.7).

These controls will reduce the information collected during the initial reconnaissance phase of attacks.
Although such security controls do not guarantee the prevention of attacks,
they are useful in limiting the attacker's prior knowledge.
For example, if attackers have less knowledge of the trained model's task, then they may fail to evade certain detection techniques for model extraction attacks~\cite{10.1007/978-3-030-62144-5_4}.

In actual development, models are often trained on publicly available datasets.
In that case, the developers cannot keep the dataset information confidential,
hence should implement other security controls.

Finally, restricting the disclosure of the system's specification may reduce the system's transparency and accountability.
Therefore, the system developers need to examine the tradeoff between security and transparency/accountability.

\section{Summary Tables and Conclusion}
\label{sec:conclusion}

We proposed the Artificial Intelligence Security Taxonomy to systematize the knowledge of threats, vulnerabilities, and security controls of ML-based systems
from the perspectives of information security and software engineering.

In \Sec{sec:overview}, we first explained the characteristics of ML-specific security,
classified the damage caused by attacks against ML-based systems, defined the notion of ML-specific security, and discussed its characteristics.
In \Sec{sec:assets:stakeholders}, we listed all relevant assets and stakeholders (addressing the motivations M1 in \Sec{sub:char1} and M2 in \Sec{sub:char2}).
In \Sec{sec:attacks:MLBS}, we provided a general taxonomy for ML-specific threats.
In \Sec{sec:V},
we collected a wide range of security controls against ML-specific threats
through an extensive review of recent literature, 
and classified the vulnerabilities and controls of an ML-based system in terms of each vulnerable asset in the system's entire lifecycle.

Throughout the paper, we emphasized that the security of ML technologies should be assessed and controlled across multiple assets in the system's entire lifecycle (corresponding to M3 in \Sec{sub:char3} and M4 in \Sec{sub:char4}).
Based on our classification, we pointed out areas of potential future research on security control techniques (M5 in \Sec{sub:char5}).

For convenience, 
we summarize the security controls to ML-specific threats
for each asset in the system lifecycle.
We show security controls to
A1, A11 (ML data sources) and A2, A12 (ML datasets)
in \Tbl{tab:control:1};
A3 (pre-trained models), A4 (learning mechanisms), and A5 (trained models)
in \Tbl{tab:control:2};
A6 (systems)
in \Tbl{tab:control:3};
A7 (data sources for system operation),
A8 (data for system operation), and
A9 (the computing environment and the operating organization)
in \Tbl{tab:control:op}.

Using these tables, developers can easily identify what security controls they can design and implement for the security of an ML-based system.
We emphasize that, since developers cannot (and need not) implement all security controls in the tables,
they should prioritize the threats and vulnerabilities of a specific system and implement security controls in order of priority.

Finally, in future work, we plan to extend our framework to other categories of machine learning,
such as unsupervised, semi-supervised, reinforcement, online, and distributed learning.

\begin{table}[t]
  \caption{List of tables about the controls of ML data sources and datasets to ML-specific threats during system development.
  }
  \label{tab:control:1}
\begin{footnotesize}
\begin{tabular}{@{}l@{}l@{}l@{}l@{}l@{}}\hlineb
\multicolumn{2}{@{}l}{\begin{tabular}{@{}l} \textgt{Assets to be} \\[-0.5ex] \textgt{controlled} \end{tabular} }  & \textgt{Threat} & \multicolumn{2}{@{}l}{\textgt{Control}} \\
\hline
\multirow{2}{*}{\begin{tabular}{@{}l} A1, \\ A11 \end{tabular}} & \multirow{2}{*}{\begin{tabular}{@{}l} ML data \\ sources \end{tabular}} & \multirow{2}{*}{\begin{tabular}{@{}l} Data poisoning \\ attack \end{tabular}} & C1.1 & \begin{tabular}{@{}l} Evaluate the trustworthiness of ML data sources \end{tabular} \\
 &  &  & C1.2 & \begin{tabular}{@{}l} Apply security controls to prevent/mitigate the poisoning \\ of ML data sources \end{tabular} \\
 &  &  & C1.3 & \begin{tabular}{@{}l} Use techniques to detect the poisoning of ML data sources \end{tabular} \\
\hline
\multirow{2}{*}{\begin{tabular}{@{}l} A2, \\ A12 \end{tabular}} & \multirow{2}{*}{\begin{tabular}{@{}l} ML \\ datasets \end{tabular}} & \multirow{2}{*}{\begin{tabular}{@{}l} Data poisoning \\ attack \end{tabular}} & C2.1 & \begin{tabular}{@{}l} Evaluate the trustworthiness of ML datasets \end{tabular} \\
 &  &  & C2.2 & \begin{tabular}{@{}l} Apply security controls to prevent/mitigate the poisoning \\ of ML datasets \end{tabular} \\
 &  &  & C2.3 & \begin{tabular}{@{}l} Use techniques to detect the poisoning of ML datasets \end{tabular} \\
 &  &  & C2.4a ~& \begin{tabular}{@{}l} Use techniques to synthesize/pre-process the ML datasets \\ to make them resilient to data poisoning \end{tabular} \\
\cline{3-5}
 &  & \begin{tabular}{@{}l} Model extraction \\ attack \end{tabular} & C2.4b & \begin{tabular}{@{}l} Use techniques to synthesize/pre-process the ML datasets \\ to mitigate model extraction \end{tabular} \\
\cline{3-5}
 &  & Evasion attack & C2.4c & \begin{tabular}{@{}l} Use techniques to synthesize/pre-process the ML datasets \\ to produce robust models against adversarial examples \end{tabular} \\
\cline{3-5}
 &  & Sponge attack & C2.4d & \begin{tabular}{@{}l} Use techniques to synthesize/pre-process the ML datasets \\ to produce robust models against sponge examples \end{tabular} \\
\cline{3-5}
 &  & \begin{tabular}{@{}l} Information \\ leakage attack \\ of training data \end{tabular} & C2.4e & \begin{tabular}{@{}l} Use techniques to synthesize/pre-process the ML datasets \\ to mitigate the leakage of sensitive information in the \\ training datasets \end{tabular} \\
\hlineb
\end{tabular}
\end{footnotesize}
\end{table}

\begin{table}[t]
  \caption{List of tables about the controls of assets to ML-specific threats during system development.
  }
  \label{tab:control:2}
\begin{footnotesize}
\begin{tabular}{@{}l@{}l@{}l@{}l@{}l@{}}\hlineb
\multicolumn{2}{@{}l}{\begin{tabular}{@{}l} \textgt{Assets to be} \\[-0.5ex] \textgt{controlled} \end{tabular} }  & \textgt{Threat} & \multicolumn{2}{@{}l}{\textgt{Control}} \\
\hline
A3 ~& \multirow{2}{*}{\begin{tabular}{@{}l} Pre-trained \\ model \end{tabular}} & \multirow{2}{*}{\begin{tabular}{@{}l}Model poisoning \\ attack \end{tabular}} & C3.1 & \begin{tabular}{@{}l} Evaluate the trustworthiness of pre-trained models \end{tabular} \\
 &  &  & C3.2 & \begin{tabular}{@{}l} Apply security controls to prevent/mitigate the \\ manipulation of pre-trained models \end{tabular} \\
 &  &  & C3.3 & \begin{tabular}{@{}l} Use techniques to detect poisoning effects from pre-trained \\ models \end{tabular} \\
 &  &  & C3.4 & \begin{tabular}{@{}l} Use techniques to remove/reduce poisoning effects from \\ pre-trained models \end{tabular} \\
\hline
A4 & \begin{tabular}{@{}l} Learning \\ mechanism \end{tabular} & \begin{tabular}{@{}l} Data poisoning \\ attack \end{tabular} & C4.1 & \begin{tabular}{@{}l} Use learning mechanisms being more resilient to data \\ poisoning \end{tabular} \\
\cline{3-5}
 &  & \multirow{2}{*}{\begin{tabular}{@{}l} Model poisoning \\ attack \end{tabular}} & C4.2 & \begin{tabular}{@{}l} Evaluate the trustworthiness of learning mechanisms \end{tabular} \\
 &  &  & C4.3 & \begin{tabular}{@{}l} Apply security controls to prevent/mitigate the \\ manipulation of learning mechanisms \end{tabular} \\
 &  &  & C4.4 & \begin{tabular}{@{}l} Use learning mechanisms that can remove/reduce \\ poisoning effects from pre-trained models \end{tabular} \\
\cline{3-5}
 &  & \begin{tabular}{@{}l} Model extraction \\ attack \end{tabular} & C4.5b ~& \begin{tabular}{@{}l} Use learning mechanisms that produce trained models \\ resilient to their extraction \end{tabular} \\
\cline{3-5}
 &  & Evasion attack & C4.5c & \begin{tabular}{@{}l} Use learning mechanisms that produce robust models \\ against adversarial examples \end{tabular} \\
\cline{3-5}
 &  & Sponge attack & C4.5d & \begin{tabular}{@{}l} Use learning mechanisms that produce robust models \\ against sponge examples \end{tabular} \\
\cline{3-5}
 &  & \begin{tabular}{@{}l} Information \\ leakage attack \\ of training data \end{tabular} & C4.5e & \begin{tabular}{@{}l} Use learning mechanisms that can prevent/mitigate \\ the leakage of sensitive information in a training dataset \end{tabular} \\
\hline
A5 & \multirow{2}{*}{\begin{tabular}{@{}l} Trained \\ model \end{tabular}} & \multirow{2}{*}{\begin{tabular}{@{}l} Model poisoning \\ attack \end{tabular}} & C5.1 & \begin{tabular}{@{}l} Apply security controls to suppress/prevent the poisoning \\ of trained models \end{tabular} \\
 &  &  & C5.2 & \begin{tabular}{@{}l} Use techniques to detect poisoning effects in trained  models \end{tabular} \\
\cline{3-5}
 &  & \begin{tabular}{@{}l} Data/model \\ poisoning attack \end{tabular} & C5.3 & \begin{tabular}{@{}l} Use techniques to remove/reduce poisoning effects from \\ trained models \end{tabular} \\
\cline{3-5}
 &  & \begin{tabular}{@{}l} Model extraction \\ attack \end{tabular} & C5.4b & \begin{tabular}{@{}l} Evaluate the risk of extraction of trained models \end{tabular} \\
\cline{3-5}
 &  & Evasion attack & C5.4c &  \begin{tabular}{@{}l} Evaluate the robustness of trained models against \\ adversarial examples \end{tabular} \\
\cline{3-5}
 &  & Sponge attack & C5.4d &  \begin{tabular}{@{}l} Evaluate the robustness of trained models against \\ sponge examples \end{tabular} \\
\cline{3-5}
 &  & \begin{tabular}{@{}l} Information \\ leakage attack \\ of training data \end{tabular} & C5.4e & \begin{tabular}{@{}l} Evaluate the risk of information leakage from trained \\ models \end{tabular} \\
\hlineb
\end{tabular}
\end{footnotesize}
\end{table}

\begin{table}[t]
  \caption{List of tables about the controls of systems to ML-specific threats during system development.
  }
  \label{tab:control:3}
\begin{footnotesize}
\begin{tabular}{@{}l@{}l@{}l@{}l@{}l@{}}\hlineb
\multicolumn{2}{@{}l}{\begin{tabular}{@{}l} \textgt{Assets to be} \\[-0.5ex] \textgt{controlled} \end{tabular} }  & \textgt{Threat} & \multicolumn{2}{@{}l}{\textgt{Control}} \\
\hline
A6.1 ~& \begin{tabular}{@{}l} Access \\ control \\ program \end{tabular} & \begin{tabular}{@{}l} Malicious input \\ of data for \\ system operation \end{tabular} & C6.1 & \begin{tabular}{@{}l} Enforce access control for ML components \\ during operation \end{tabular} \\
\hline
A6.2 & \begin{tabular}{@{}l} Pre-processing \\ program \end{tabular} & \begin{tabular}{@{}l} Malicious input \\ of data for \\ system operation \end{tabular} & C6.2 & \begin{tabular}{@{}l} Use techniques to detect/pre-process/restrict \\ malicious input to ML components during \\ operation \end{tabular} \\
\hline
A6.3 & \begin{tabular}{@{}l} ML component \end{tabular} & \begin{tabular}{@{}l} Exploitation of \\ poisoned models \end{tabular} & \begin{tabular}{@{}l} C1 \\ to C5 \end{tabular} & \begin{tabular}{@{}l} Apply the security controls of assets A1 to \\ A5 against poisoning attacks \\ (See \Tbls{tab:control:1} and \ref{tab:control:2}) \end{tabular} \\
\cline{3-5}
 &  & \multirow{2}{*}{\begin{tabular}{@{}l} Model extraction \\ attack \end{tabular}} & C5.5b & \multirow{3}{*}{\begin{tabular}{@{}l} Use techniques to improve trained models \\ to mitigate the leakage of information on \\ trained models \end{tabular}} \\
\\
\\
\cline{3-5}
 &  & \begin{tabular}{@{}l} Evasion attack \end{tabular} & C5.5c & \begin{tabular}{@{}l} Use techniques to improve the robustness \\ of trained models against adversarial examples \end{tabular} \\
\cline{3-5}
 &  & \begin{tabular}{@{}l} Sponge attack \end{tabular} & C5.5d & \begin{tabular}{@{}l} Use techniques to improve the robustness \\ of trained models against sponge examples \end{tabular} \\
\cline{3-5}
 &  & \begin{tabular}{@{}l} Information \\ leakage attack \\ of training data \end{tabular} & C5.5e & \begin{tabular}{@{}l} Use techniques to improve trained models \\ to mitigate the leakage of information in the \\ training dataset \end{tabular} \\
\hline
A6.4 & \begin{tabular}{@{}l} Post-processing \\ program \end{tabular} & \begin{tabular}{@{}l} Malicious input \\ of data for \\ system operation \end{tabular} & C6.4 & \begin{tabular}{@{}l} Restrict the disclosure of the output and \\ internal information of ML components during \\ operation \end{tabular} \\
\hline
A6.5 & \begin{tabular}{@{}l} Monitoring/risk \\ treatment \\ program \end{tabular} & \begin{tabular}{@{}l} All types of \\ ML-specific attacks \end{tabular} & C6.5 & \begin{tabular}{@{}l} Apply security controls to monitor the \\ system's behavior and treat the risks caused \\ by ML components \end{tabular} \\
\hline
\begin{tabular}{@{}l} A6.1 \\ to A6.5 \end{tabular} & \begin{tabular}{@{}l} Programs directly \\ supporting the \\ model operation \end{tabular} &  \begin{tabular}{@{}l} Conventional threats \\ to systems \end{tabular} & C6.6 & \begin{tabular}{@{}l} Apply security controls for the vulnerability \\ of conventional software \end{tabular} \\
\hline
A6.6 & \begin{tabular}{@{}l} Other \\ conventional \\ software \\ components \end{tabular} & \begin{tabular}{@{}l} Conventional threats \\ to systems \end{tabular} & C6.6 & \begin{tabular}{@{}l} Apply security controls for the vulnerability \\ of conventional software \end{tabular} \\
\hline
A6.7 & \begin{tabular}{@{}l} System \\ specification, etc. \end{tabular} & \begin{tabular}{@{}l} All types of \\ ML-specific attacks \end{tabular} & C6.7 & \begin{tabular}{@{}l} Restrict the disclosure of the ML datasets, \\ the trained models, the other system \\ specifications, and their related information \end{tabular} \\
\hlineb
\end{tabular}
\end{footnotesize}
\end{table}

\begin{table}[t]
  \caption{List of tables about the controls to ML-specific threats during system operation.
  }
  \label{tab:control:op}
\begin{footnotesize}
\begin{tabular}{@{}l@{}l@{}l@{}l@{}l@{}}\hlineb
\multicolumn{2}{@{}l}{\begin{tabular}{@{}l} \textgt{Assets to be} \\[-0.5ex] \textgt{controlled} \end{tabular} }  & \textgt{Threat} & \textgt{Control} & \textgt{} \\
\hline
A7 ~& \multirow{4}{*}{\begin{tabular}{@{}l} Source of \\ data for \\ system \\ operation \end{tabular}} & \multirow{3}{*}{\begin{tabular}{@{}l} Malicious input \\ of data for \\ system operation \end{tabular}} & C7.1 & \begin{tabular}{@{}l} Evaluate the trustworthiness of sources of data for \\ system operation \end{tabular} \\
 &  &  & C7.2 & \begin{tabular}{@{}l} Apply security controls to prevent/mitigate the \\ manipulation of sources of data for system operation \end{tabular} \\
 &  &  & C7.3 & \begin{tabular}{@{}l} Use techniques to detect the manipulation of sources \\ of data for system operation \end{tabular} \\
\hline
A8 & \multirow{3}{*}{\begin{tabular}{@{}l} Data for \\ system \\ operation \end{tabular}} & \multirow{3}{*}{\begin{tabular}{@{}l} Malicious input \\ of data for \\ system operation \end{tabular}} & C8.1 & \begin{tabular}{@{}l} Evaluate the trustworthiness of data for system operation \end{tabular} \\
 &  &  & C8.2 & \begin{tabular}{@{}l} Apply security controls to prevent/mitigate the \\ manipulation of data for system operation \end{tabular} \\
 &  &  & C8.3 & \begin{tabular}{@{}l} Use techniques to detect the manipulation of data for \\ system operation \end{tabular} \\
\hline
A9 & \multirow{6}{*}{\begin{tabular}{@{}l} Computing \\ environment \\ \& operating \\ organization \\ during system \\ operation \end{tabular}} & \multirow{2}{*}{\begin{tabular}{@{}l} All types \\ of threats \end{tabular}} & C9.1 & \begin{tabular}{@{}l} Apply security controls for the vulnerability of the \\ computing environment and the operating organization \\ during system operation \end{tabular} \\
 &  &  & C9.2 & \begin{tabular}{@{}l} Update security controls continuously by the system \\ operator to cope with the changes in the system and \\ the environment \end{tabular} \\
 &  &  & C9.3 & \begin{tabular}{@{}l} Enable the system operator to monitor attacks and \\ damage manually \end{tabular} \\
\hlineb
\end{tabular}
\end{footnotesize}
\end{table}

\backmatter

\bmhead{Acknowledgments}
This paper is based on results obtained from a project,
JPNP20006, commissioned by the New Energy and Industrial Technology Development
Organization (NEDO).
Yusuke Kawamoto is supported by JST, PRESTO Grant Number JPMJPR2022, Japan.

\bibliography{%
ref-survey,
ref10.1-r1,
ref10.2,
ref10.3.2,
ref10.3.5,
ref10.3.6,
ref10.3.7,
ref10.5-r2,
ref-sponge,
refothers-r1}

\end{document}